\documentclass[twocolumn]{aastex63}

\usepackage[utf8]{inputenc}
\usepackage{hyperref}
\usepackage{amsmath}

\usepackage{silence}
\usepackage{graphicx}

\usepackage{xcolor}

\usepackage{float}
\usepackage{color,soul}
\usepackage{bm}

\usepackage{wasysym}
\usepackage{upgreek}
\usepackage{natbib}
\usepackage{makecell}
\usepackage{enumitem}
\usepackage{multirow}
\usepackage{xspace}
\usepackage{todonotes}

\usepackage{etoolbox}

\makeatletter

\patchcmd{\NAT@citex}
  {\@citea\NAT@hyper@{%
     \NAT@nmfmt{\NAT@nm}%
     \hyper@natlinkbreak{\NAT@aysep\NAT@spacechar}{\@citeb\@extra@b@citeb}%
     \NAT@date}}
  {\@citea\NAT@nmfmt{\NAT@nm}%
   \NAT@aysep\NAT@spacechar\NAT@hyper@{\NAT@date}}{}{}

\patchcmd{\NAT@citex}
  {\@citea\NAT@hyper@{%
     \NAT@nmfmt{\NAT@nm}%
     \hyper@natlinkbreak{\NAT@spacechar\NAT@@open\if*#1*\else#1\NAT@spacechar\fi}%
       {\@citeb\@extra@b@citeb}%
     \NAT@date}}
  {\@citea\NAT@nmfmt{\NAT@nm}%
   \NAT@spacechar\NAT@@open\if*#1*\else#1\NAT@spacechar\fi\NAT@hyper@{\NAT@date}}
  {}{}

\makeatother

\shorttitle{CASPAR}
\shortauthors{Betti et al.}

\begin{document}

\title{The Comprehensive Archive of Substellar and Planetary Accretion Rates}

\author[0000-0002-8667-6428]{S. K. Betti}
\affiliation{Department of Astronomy, University of Massachusetts, Amherst, MA 01003, USA}

\author[0000-0002-7821-0695]{K. B. Follette}
\affiliation{Department of Physics and Astronomy, Amherst College, Amherst, MA 01003, USA}

\author[0000-0002-4479-8291]{K. Ward-Duong}
\affiliation{Department of Astronomy, Smith College, Northampton, MA, 01063, USA}

\author[0000-0003-2461-6881]{A. E. Peck}
\affiliation{Department of Astronomy, New Mexico State University, Las Cruces, NM 88003}
\affiliation{Department of Astronomy, Smith College, Northampton, MA, 01063, USA}

\author[0000-0003-0568-9225]{Y. Aoyama}
\affiliation{Kavli Institute for Astronomy and Astrophysics, Peking University, Beijing 100871, P.R. China}

\author[0000-0001-8642-5867]{J. Bary}
\affiliation{Department of Physics and Astronomy, Colgate University, Hamilton, NY, 13346, USA}

\author{B. Dacus}
\affiliation{Department of Astronomy and Astrophysics, University of California, San Diego,  La Jolla, CA 92093, USA}
\affiliation{Department of Physics and Astronomy, Amherst College, Amherst, MA 01003, USA}

\author[0000-0002-3232-665X]{S. Edwards}
\affiliation{Department of Astronomy, Smith College, Northampton, MA, 01063, USA}

\author[0000-0002-2919-7500]{G.-D. Marleau}
\affiliation{Fakult\"at f\"ur Physik, Universit\"at Duisburg-Essen, Lotharstra\ss{}e 1, 47057 Duisburg, Germany}
\affiliation{Institut f\"ur Astronomie und Astrophysik, Universit\"at T\"ubingen, Auf der Morgenstelle 10, 72076 T\"ubingen, Germany}
\affiliation{Physikalisches Institut, Universit\"at Bern, Gesellschaftsstr.~6, 3012 Bern, Switzerland }
\affiliation{Max-Planck-Institut f\"ur Astronomie, K\"onigstuhl 17, 69117 Heidelberg, Germany}

\author[0000-0001-9502-3448]{K. Mohamed}
\affiliation{Department of Astronomy, Boston University, Boston, MA 02215}
\affiliation{Department of Physics and Astronomy, Amherst College, Amherst, MA 01003, USA}

\author{J. Palmo}
\affiliation{Department of Civil and Environmental Engineering, Massachusetts Institute of Technology, Cambridge, MA 02139}
\affiliation{Department of Physics and Astronomy, Amherst College, Amherst, MA 01003, USA}

\author[0000-0002-1144-6708]{C. Plunkett}
\affiliation{Department of Physics and Astronomy, Amherst College, Amherst, MA 01003, USA}

\author[0000-0003-1639-510X]{C. Robinson}
\affiliation{Department of Physics and Astronomy, Amherst College, Amherst, MA 01003, USA}

\author{H. Wang}
\affiliation{Department of Physics and Astronomy, Amherst College, Amherst, MA 01003, USA}

\correspondingauthor{S. K. Betti}
\email{sbetti@stsci.edu}

\begin{abstract}
Accretion rates ($\dot{M}$) of young stars show a strong correlation with object mass ($M$); however, extension of the $\dot{M}-M$ relation into the substellar regime is less certain.  
Here, we present the Comprehensive Archive of Substellar and Planetary Accretion Rates (CASPAR), the largest to-date compilation of substellar accretion diagnostics. CASPAR includes: 658 stars, 130 brown dwarfs, and 10 bound planetary mass companions. In this work, we investigate the contribution of methodological systematics to scatter in the $\dot{M}-M$ relation, and compare brown dwarfs to stars.
In our analysis, we rederive all quantities using self-consistent models, distances, and empirical line flux to accretion luminosity scaling relations to reduce methodological systematics. This treatment decreases the original $1\sigma$ scatter in the $\log \dot{M}-\log M$ relation by $\sim17$\%, suggesting that it makes only a small contribution to the dispersion. CASPAR rederived values are best fit by $\dot{M}\propto M^{2.02\pm0.06}$ from 10~$M_\mathrm{J}$ to 2~$M_\odot$, confirming previous results. 
However, we argue that the brown dwarf and stellar populations are better described separately and by accounting for both mass and age. Therefore, we derive separate age-dependent $\dot{M}-M$ relations for these regions, and find a steepening in the brown dwarf $\dot{M}-M$ slope with age.   Within this mass regime, the scatter decreases from 1.36 dex to 0.94 dex, a change of $\sim$44\%. This result highlights the significant role that evolution plays in the overall spread of accretion rates, and suggests that brown dwarfs evolve faster than stars, potentially as a result of different accretion mechanisms.
\end{abstract}

\section{Introduction} \label{sec:intro}
In the classical picture of star formation, molecular cloud cores collapse under gravity to form new stars. As the cores collapse, rotational velocity and conservation of angular momentum causes infalling material to settle into a circumstellar disk \citep{Hartmann1998}.  These primordial disks have been found to have a lifetime of $\sim$1--10 Myr \citep[e.g.][]{Strom1989, Armitage2003, Sicilia-Aguilar2006, Li2016}, during which time they provide the material essential for both planet formation \citep{Mamajek2009} and stellar accretion \citep{Hartmann2016}. The evolution and dispersal of the disk unfolds through several processes.  Planet formation occurs through core accretion of planetesimals in the inner few astronomical units \citep{Safronov1969, Hayashi1985, Pollack1996} leading to terrestrial planet formation, while in the outer disk, core accretion, fragmentation, and instabilities are theorized to form giant gaseous protoplanets \citep{Kuiper1951, Cameron1978, Boss1997, Bate2002, Bate2003, Johnson2003, Rafikov2005, Cai2006}. Through the T~Tauri phase, the disks are also actively accreting material onto the star, enabled by a combination of viscous accretion through the magnetorotational instability \citep{BalbusHawley1991} and/or MHD disk winds, depending on physical conditions in the disk \citep{Pascucci2023, Lesur2023}. 
Additionally strong near-UV, far-UV, and/or X-ray radiation from the star and its accretion shock can heat the gaseous disk surface leading to thermally driven photoevaporative winds beyond the gravitational radius, which likely account for the final clearing of the disk gas  \citep[][and references therein]{Alexander2014}. 

Within a few stellar radii of the star \citep[traditionally assumed to be $\sim 5\ R_\odot$;][]{Gullbring1998}, the disk is interrupted by strong stellar magnetic fields and disk material flows to the stellar surface along accretion columns following magnetic field lines resulting in strong shocks on the stellar surface. Resultant emission from the accretion onto the star includes broad emission lines in the free-falling magnetospheric flows \citep{Muzerolle2001, Hartmann2016}, though \citet{Dupree2014} suggested that in the case of hydrogen, the broad lines could be formed in a turbulent postshock region, and forbidden lines from accretion shocks and winds \citep{Hartigan1995}. When the gas shocks on the stellar photosphere, the already fully ionized gas heats to $10^6$~K. The optically thin preshock region is seen primarily as Balmer continuum excess \citep{Valenti1993, Gullbring1998, Gullbring2000, Calvet1998}, while the optically thick post-shock region emits Paschen continuum excess\footnote{In the near infrared (NIR), excess continuum emission from dust is also produced at the inner edge of the disk due to heating by radiation from the photosphere and shocked regions \citep{Johns-Krull2001,Muzerolle2003, Fischer2011}.} \citep{Calvet1998}. These sources of excess continuum emission result in veiling of photospheric absorption lines. 
Accretion rates measured from both optically thin and thick shock regions can be inferred from the total excess luminosity produced by accretion; however, there is currently no direct method to measure the mass accretion rate ($\dot M$) from the emission produced in the accretion flows. Instead, a scaling relation must be applied to relate a single emission line luminosity to a mass accretion rate. 

Comprehensive multiwavelength studies have found that $\dot M$ decreases with decreasing stellar mass ($M$) \citep{Muzerolle2003, Calvet2004, Herczeg2008, Alcala2014, Manara2017}, following a power law of $\dot M \propto M^{2}$ in the stellar regime.
This mass accretion rate$-$mass ($\dot M-M$) relation has been assumed to extend from the stellar to the substellar ($M \leq 0.075M_{\odot}$) regime with no variation in slope \citep[e.g.,][]{Muzerolle2003, Mohanty2005, Muzerolle2005}, though some studies suggest a break to a steeper relation around 0.2~$M_\odot$ for older star forming regions \citep{Alcala2017, Manara2017b}.  Additionally, at all masses, there is significant 1--2 dex scatter in accretion rates. Within the stellar regime, \citet{Manara2022} assert that the majority of this scatter results from physical variation and not observational uncertainty.   

Various studies have looked at possible physical mechanisms responsible for this dispersion in the $\dot M-M$ relation. These include: intrinsic variability and the decrease of $\dot M$ with age \citep[e.g.][]{Natta2006, Costigan2014, Venuti2014, Hartmann2016}, differences in the properties of star-forming cores \citep[e.g.][]{Dullemond2006, Clarke2006, Ercolano2014}, competition among accretion mechanisms \citep[viscosity or gravitational instability, e.g.][]{Vorobyov2008, Vorobyov2009, DeSouza2017}, multiplicity \citep[e.g.][]{Zagaria2022} and, for PMCs, differences in the instability threshold and reservoir for accretion as a result of disk fragmentation \citep[e.g.][]{Stamatellos2015}. However, systematic studies of large numbers of accreting \textit{substellar} objects are lacking, and it is not clear if this proposed explanation holds in this low mass regime.  

Additionally, scaling relationships between line emission and accretion luminosity have been empirically developed and calibrated for stars for a wide variety of emission lines \citep{Alcala2014, Alcala2017, Rigliaco2011, Natta2004}. It is not clear to what extent empirical scaling relations are valid in the substellar regime, where potential differences in accretion (magnetospheric, planetary shock) and physical parameters (energy loss, magnetic field strength, temperature of accreting gas, gravitational potential, disk mass), could alter the relationship between line luminosity and mass accretion rate \citep{Thanathibodee2019, Aoyama2020}. In order to understand the origin and accretion of both bound planetary mass companions (PMCs; which include protoplanet candidates \citep[e.g., PDS 70b and c, Delorme 1 (AB)b, LkCa 15b;][]{Haffert2019, Wagner2018, Sallum2015, Eriksson2020, Ringqvist2023, Betti2022b} and brown dwarf (BD) companions, objects in bound orbits around a higher mass host, \citep[which we define as bound objects below $M<30\ M_\mathrm{J}$; c.f.][]{Martinez2019}), and young BDs (hereafter, all considered substellar objects), we first must characterize the physical (e.g., variability, age), and systematic (e.g., accretion rate tracer, evolutionary model) properties that affect accretion rate estimates.  

The aim of this paper is threefold: a) to provide the largest compilation to-date of brown dwarf and protoplanet accretion rates derived under a uniform methodology, b) investigate methodological differences in the scatter of the $\dot M-M$ relation between stars and brown dwarfs, and c) determine whether the statistics of $\dot M$ measurements suggest accretion differences between these mass populations.
In Section~\ref{sec:database}, we give an overview of the Comprehensive Archive of Substellar and Planetary Accretion Rates (CASPAR). In Section~\ref{sec:rederive}, we rederive object properties (e.g. mass, distance, temperature) in a consistent manner.  In Section~\ref{sec:linfit}, we detail the technique we applied to derive linear fits for the $\dot M-M$ relation.  We also present an updated $\dot M-M$ relation and discuss the role of methodology in producing scatter in Section~\ref{sec:scatter}.  In Section~\ref{sec:physicalscatter}, we quantify the contribution of various drivers producing the physical scatter observed in the overall $\dot M-M$ relation, and in Section~\ref{sec:BDpop}, we focus on the BD population. In Section~\ref{sec:discuss}, we discuss how these phenomena affect our interpretation of accretion in the substellar regime. Results are then summarized in Section~\ref{sec:conclusion}.

\section{Overview of the Database}\label{sec:database}
We have assembled accretion rates for young planetary mass companions, brown dwarfs, and Classical T Tauri stars (CTTS) from large surveys of accreting objects, as well as individual object papers.  This database consists of two parts: a compilation of published accretion properties, unmodified from their source publications (hereafter the Literature Database), and a unified re-derivation of accretion properties from these studies, CASPAR\footnote{CASPAR is openly available on Zenodo:\\ \url{https://doi.org/10.5281/zenodo.8393054}.}.

\begin{figure*}[hpt!]
    \centering
    \epsscale{1.1}
    \plotone{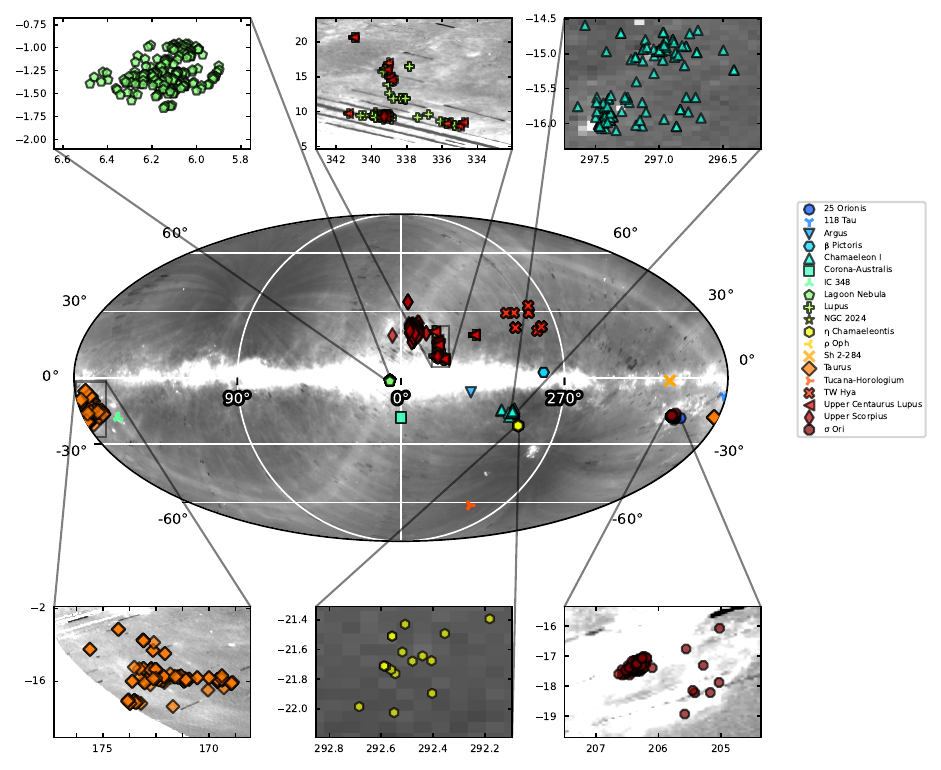}
    \caption{All-sky map indicating all objects in CASPAR colored by star-forming region or association overlaid on a 65~$\mu$m all-sky map \citep{Doi2015,Takita2015}.  \textit{Insets:} Enlarged views of several nearby star forming regions. }
    \label{fig1:skymap}
\end{figure*}

We have focused on collecting properties for known accreting substellar objects. Within the Literature Database, 86 objects are considered substellar \citep[below the hydrogen burning limit, $M<0.075\ M_\odot$, e.g.][]{Mohanty2005}, of which 10 are PMCs (5 are protoplanets, 5 are $M< 30\ M_\mathrm{J}$ brown dwarfs). The database also includes a substantial compilation of CTTS accretion rates for stars later than G spectral type. 
We exclude Herbig stars, as detailed accretion census papers for this population already exist \citep[e.g.,][]{Guzman-Diaz2021,Vioque2022} and we are particularly focused on substellar accretion.  To-date, we have compiled data for 798 objects from 46 studies for a total of 1058 independent accretion measurements spanning 24 years, from 1998 to early 2022.  The list of references is given in Table~\ref{tab:litrefs1}. Sky positions for all objects are shown in Fig.~\ref{fig1:skymap}.  As many of the objects are in associations and clusters with small angular scales, they appear as a single point.  We show zoomed-in views of six of the regions as insets. 

\begin{deluxetable*}{llc}					
\tablecaption{Literature Reference Star Forming Regions \label{tab:litrefs1}}		
\tablewidth{0pt}
\tablehead{\colhead{Reference} & \colhead{Star Forming region} & \colhead{\# Objects}	}				
\startdata					
\citet{Alcala2014}	&	Lupus	&	36	\\
\citet{Alcala2017}	&	Lupus	&	43	\\
\citet{Alcala2019}	&	Lupus 4	&	1	\\
\citet{Alcala2020}	&	Lupus	&	1	\\
\citet{Alcala2021}	&	Taurus-Auriga	&	5	\\
\citet{Betti2022b}	&	Tucana-Horologium	&	1	\\
\citet{Bowler2011}	&	Upper Scorpius	&	1	\\
\citet{Close2014}	&	Sco OB2-2	&	1	\\
\citet{Comeron2010}	&	$\rho$ Ophiuchus	&	1	\\
\citet{Eriksson2020}	&	Tucana-Horologium	&	1	\\
\citet{Espaillat2008}	&	25 Orionis	&	1	\\
\citet{Gatti2006}	&	$\rho$ Ophiuchus	&	16	\\
\citet{Gatti2008}	&	$\sigma$ Orionis	&	35	\\
\citet{Gullbring1998}, \citet{Calvet1998}	&	Taurus	&	17	\\
\citet{Haffert2019}	&	Centaurus	&	2	\\
\citet{Hashimoto2020}	&	Centaurus	&	2	\\
\citet{Herczeg2008}	&	Taurus, TW HyA	&	24	\\
\citet{Herczeg2009}	&	Upper Scorpius, TW HyA	&	12	\\
\citet{Ingleby2013}	&	Taurus, Chamaeleon I	&	19	\\
\citet{Kalari2015}	&	Lagoon Nebula	&	225	\\
\citet{Kalari2015b}	&	Sh 2-284	&	3	\\
\citet{Lee2020}	&	Argus	&	1	\\
\citet{Manara2015}	&	$\rho$ Ophiuchus	&	17	\\
\citet{Manara2016}	&	Chamaeleon I	&	38	\\
\citet{Manara2017}	&	Chamaeleon I	&	49	\\
\citet{Manara2020}	&	Upper Scorpius	&	35	\\
\citet{Manara2021}	&	Orion OB1	&	11	\\
\citet{Mohanty2005}	&	Chamaeleon I, IC348, Taurus	&	22	\\
\citet{Muzerolle2003}	&	Taurus, IC348	&	13	\\
\citet{Muzerolle2005}	&	Taurus, Chamaeleon I	&	33	\\
\citet{Natta2004}	&	Chamaeleon I, $\rho$ Ophiuchus	&	19	\\
\citet{Natta2006}	&	$\rho$ Ophiuchus	&	112	\\
\citet{Nguyen-Thanh2020}	&	Upper Scorpius	&	1	\\
\citet{Petrus2020}	&	Upper Scorpius	&	7	\\
\citet{Pouilly2020}	&	Taurus	&	1	\\
\citet{Rigliaco2011}	&	$\sigma$ Orionis	&	63	\\
\citet{Rigliaco2012}	&	$\sigma$ Orionis	&	8	\\
\citet{Rugel2018}	&	$\eta$ Chamaeleontis	&	15	\\
\citet{Sallum2015}	&		&	1	\\
\citet{Salyk2013}	&	Taurus, Upper Centaurus-Lupus, 	&	35	\\
	&	$\rho$ Ophiuchus,  Upper Scorpius,	&		\\
	&	Lupus, Chamaeleon I	&		\\
\citet{Santamaria-Miranda2018}	&	$\rho$ Ophiuchus	&	1	\\
\citet{Venuti2019}	&	TW Hydrae	&	9	\\
\citet{Wagner2018}	&	Upper Centaurus Lupus	&	1	\\
\citet{White2003}	&	Taurus-Auriga	&	10	\\
\citet{Wu2015}	&	Chamaeleon I	&	1	\\
\citet{Wu2017}	&	Lupus	&	1	\\
\citet{Zhou2014}	&	Taurus, Lupus, Upper Scorpius	&	3	\\
\enddata
\end{deluxetable*}	

\begin{figure}[hpt!]
    \centering
    \begin{minipage}[c]{\linewidth}
        \includegraphics[width=\linewidth]{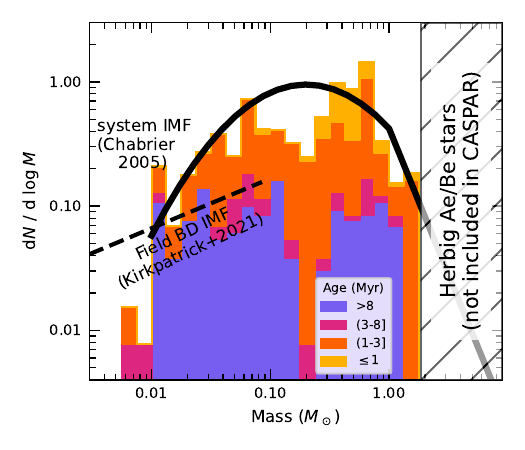}
        \caption{Mass function (d$N$/d$\log M$) of the 793 objects in CASPAR colored by stellar age (stacked histogram). The initial mass function (IMF) for multiple systems from \citet{Chabrier2005} and the field BD IMF from \citet{Kirkpatrick2021} are shown as a comparison.  Herbig Ae/Be stars are purposely excluded from CASPAR as we are primarily focused on substellar objects. }   
        \label{complete}
    \end{minipage}
    \begin{minipage}[c]{\linewidth}
        \includegraphics[width=\linewidth]{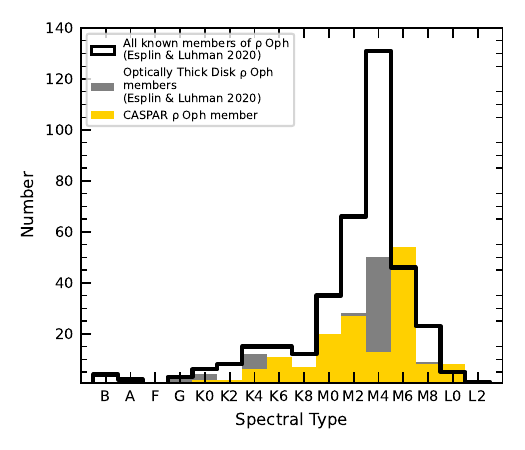}
        \caption{Distribution of spectral types in $\rho$ Ophiuchus from the complete census survey (black), and systems with optically thick disks (gray) from \citet[][binned as in their Fig.~22]{Esplin2020}.  Overlaid in yellow are the CASPAR $\rho$ Ophiuchus objects.  } 
        \label{tauruscomplete} 
    \end{minipage}
\end{figure}

As this sample is compiled from many individual studies, it is an incomplete survey of nearby objects in both mass and volume.  In Fig.~\ref{complete}, we show the mass function of our sample with uniformly rederived masses (as discussed in Sec.~\ref{sec:rederive}), colored by age and compared to the \citet{Chabrier2005} initial mass function (IMF) and \citet{Kirkpatrick2021} field brown dwarf mass function, all normalized so the integral over mass is one. The mass distribution of objects in CASPAR is consistent with the IMF, though there is a difference at $0.1-0.3\ M_\odot$, likely due to undersampling of objects at all ages in this mass range. A majority of CASPAR objects are young ($<3$ Myr), as expected since disk fraction rapidly declines after $2.5-3$ Myr \citep{Mamajek2009}.  

\begin{deluxetable*}{ll}
\tablecaption{CASPAR sections$^*$\label{tab:caspar sections}}
\tablewidth{0pt}
\tablehead{\colhead{Section} & \colhead{Description}}
\startdata
ID information & source and literature reference IDs \\
flags & duplication, binary, companion flags \\
kinematic, photometry, \& age &  6D Gaia kinematics, NIR photometry, and age/associations \\
reference \& physical parameters & literature observations information and stellar information\\
emission lines & individual emission line flux and accretion rates\\
accretion rates & final accretion luminosity and accretion rates\\
model \& scaling references & references for spectral type/temperature conversions, evolutionary models, and scaling relations
\enddata
\tablenotetext{*}{See Table~\ref{tab:CASPAR} for description of individual columns} 
\end{deluxetable*} 

From the \citet{Chabrier2005} IMF, if all of the star forming regions follow the same IMF, we would expect $\sim20\%$ more substellar objects.  These missing objects have either a) not been surveyed or b) were initially missed when compiling CASPAR. For example, when we compare CASPAR objects in $\rho$ Ophiuchus to the census from \citet{Esplin2020} (complete up to spectral type earlier than M6), we find (as shown in Fig.~\ref{tauruscomplete}) that CASPAR includes $86\%$ of all known $\rho$ Ophiuchus substellar objects (32/37; $M<0.075\ M_\odot; \leq \mathrm{M}5.5$) with optically thick disks (which we use as a proxy for potential to be accreting). 
 
Broad classes of object properties included in the database are summarized in Table~\ref{tab:caspar sections}, with individual column headers listed in Appendix~\ref{app:CASPAR}. Both the Literature Database and CASPAR have identical columns.  Each accretion rate is assigned its own row and a unique number identifier, identical between the two databases.  Therefore, an object observed at multiple epochs has multiple unique number identifiers.  Each object is also identified with a unique name.  Mass accretion rates have been measured from four broad accretion diagnostics families, namely: continuum excess, line luminosity, H$\alpha$ photometric luminosity, and line profile. In Appendix~\ref{extra}, we discuss in more detail the process of compiling the literature database, kinematic, photometric, and age information for each object. 

We show all literature database accretion rates as a function of mass colored by their accretion diagnostic in Fig.~\ref{fig:origIndiv_lines_mmdot}. Overall, we see that the accretion rates vary by 2 orders of magnitude while still following an empirical power law relationship between accretion rate and mass, $\dot{M}\propto M^{2.15}$ (black line) similar to other empirically derived relations \citep{Muzerolle2003, Muzerolle2005, Mohanty2005, Alcala2017}.  However, as shown in the botton panel of Fig.~\ref{fig:origIndiv_lines_mmdot}, we do find systematic offsets and variable slopes in the $\dot{M}\propto M^x$ relation when we fit by accretion diagnostic (see Section~\ref{sec:linfit} for fit details), which will be discussed in Section~\ref{acc diag systematics}.

\section{Unified derivation of quantities}\label{sec:rederive}
\begin{figure*}[tb]
    \centering
    \includegraphics[width=0.8\linewidth]{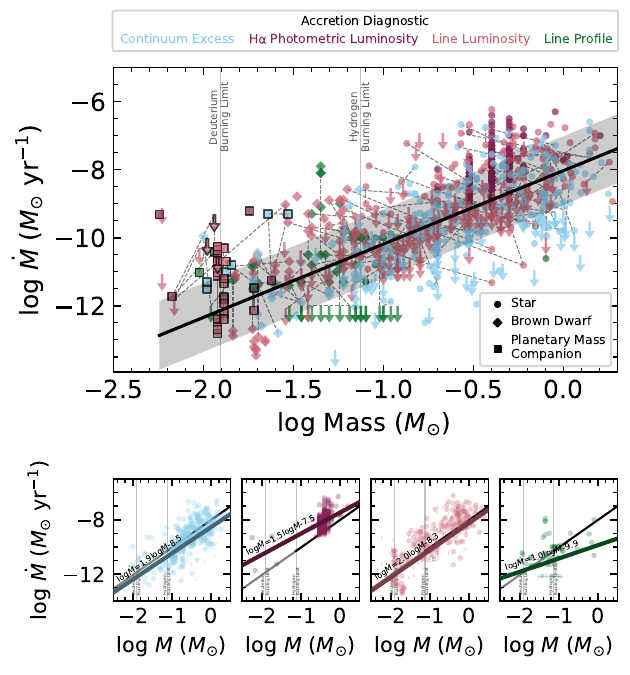}
    \caption{Literature Database of accreting objects. 
     \textit{top}: Accretion rate vs mass colored by accretion diagnostic.  Dashed gray lines indicate objects with multiple accretion rate and mass estimates.   The best fit relation, $\dot M \sim M^{2.16}$ is shown with a black solid line, with the 1$\sigma$ upper limit in gray.  Bound PMCs are indicated by thicker-edged squares, while stars are indicated by circles and brown dwarfs by diamonds. Upper limits are shown by downward arrows.  
     \textit{bottom}: Accretion rate vs mass for separate accretion diagnostics: continuum excess (light blue), H$\alpha$ photometric luminosity (dark red), line luminosity (salmon), and line profile (green).  The solid lines show the best fit for each diagnostic with the black/gray line indicating the overal fit from the top panel. 
     The vertical gray lines indicate the hydrogen and deuterium burning limits.
     Overall, the literature database follows similar trends to previous surveys while highlighting the strong scatter and variation in accretion rate and mass resulting from different methodologies. }
     \label{fig:origIndiv_lines_mmdot}
\end{figure*}

 The studies in the literature compilation come from a variety of instruments, analysis pipelines, and accretion tracers, which likely contributes to the wide dispersion of accretion rates at a single mass (i.e., the scatter).  Additionally, the dependence of mass and radius estimates on the application of a variety of different evolutionary models and spectral fitting tools also introduces scatter (see gray dashed lines in the top panel of Fig.~\ref{fig:origIndiv_lines_mmdot}). In order to remove these effects, we first investigate the dispersion introduced by methodology by re-estimating object and accretion parameters under a unified set of assumptions and by comparing them to Literature Database values. Estimates of PMC spectral types and masses are highly uncertain and have been estimated from a variety of methods including: kinematics, orbital fitting, and spectral fitting. Due to the larger uncertainties in deriving accurate masses for the lowest mass objects, we focus here only on stars and BDs. Updating the PMC entries so that their masses are uniform and comparable to the full CASPAR sample will be the subject of future work.  As a result, in the remainder of this study, we use the Literature Database mass and accretion rate estimates for PMCs in our plots and calculations.  While we fit this PMC population and report it here, we do not attempt to infer any trends, as these accretion rates and masses are not rederived under a unified system.

In this work we update and unify literature values by performing the following modifications:
\begin{itemize}
    \item adopt Gaia DR2/EDR3 \citep{GAIADR22018, GAIAEDR32021} distances when available \citep[N=604, ][]{Bailer-Jones2018, Bailer-Jones2021} 
    \item adopt single ages for each star forming region (or subregion, where available; see Section~\ref{sec:kin_phot_age}),
    \item adopt the spectral type - temperature conversion from \citet{Herczeg2014}
    \item extract mass, luminosity, radius, and surface gravity using the MIST MESA models \citep{Paxton2011, Paxton2013, Paxton2015, Dotter2016, Choi2016} K spectral types, and \citet{Baraffe2015} evolutionary models for all others
    \item calculate accretion luminosities using the \citet{Alcala2017} $L_\mathrm{acc}-L_\mathrm{line}$ scaling relationships.  For excess continuum based accretion luminosity estimates, we scale by $d^2$ from Gaia DR2/EDR3. 
    \item From accretion luminosity, calculate the mass accretion rate as
\begin{equation}\label{eqn1}
    \dot{M} = \left(1-\frac{R}{R_\mathrm{in}}\right)^{-1} \frac{L_\mathrm{acc} R}{GM},
\end{equation}
where $\dot M$ is the mass accretion rate, $R$ is the stellar radius, $R_\mathrm{in}$ is the truncation radius, which we assume to be $5\ R_\odot$ \citep{Gullbring1998}, $M$ is the stellar mass, $G$ is the gravitational constant, and $L_\mathrm{acc}$ is the accretion luminosity. 
\end{itemize}
We describe the unified methodology for deriving each parameter in full detail in Appendix~\ref{app:rederivephysical}.

\begin{figure*}[htp!]
    \centering
    \includegraphics[width=\linewidth]{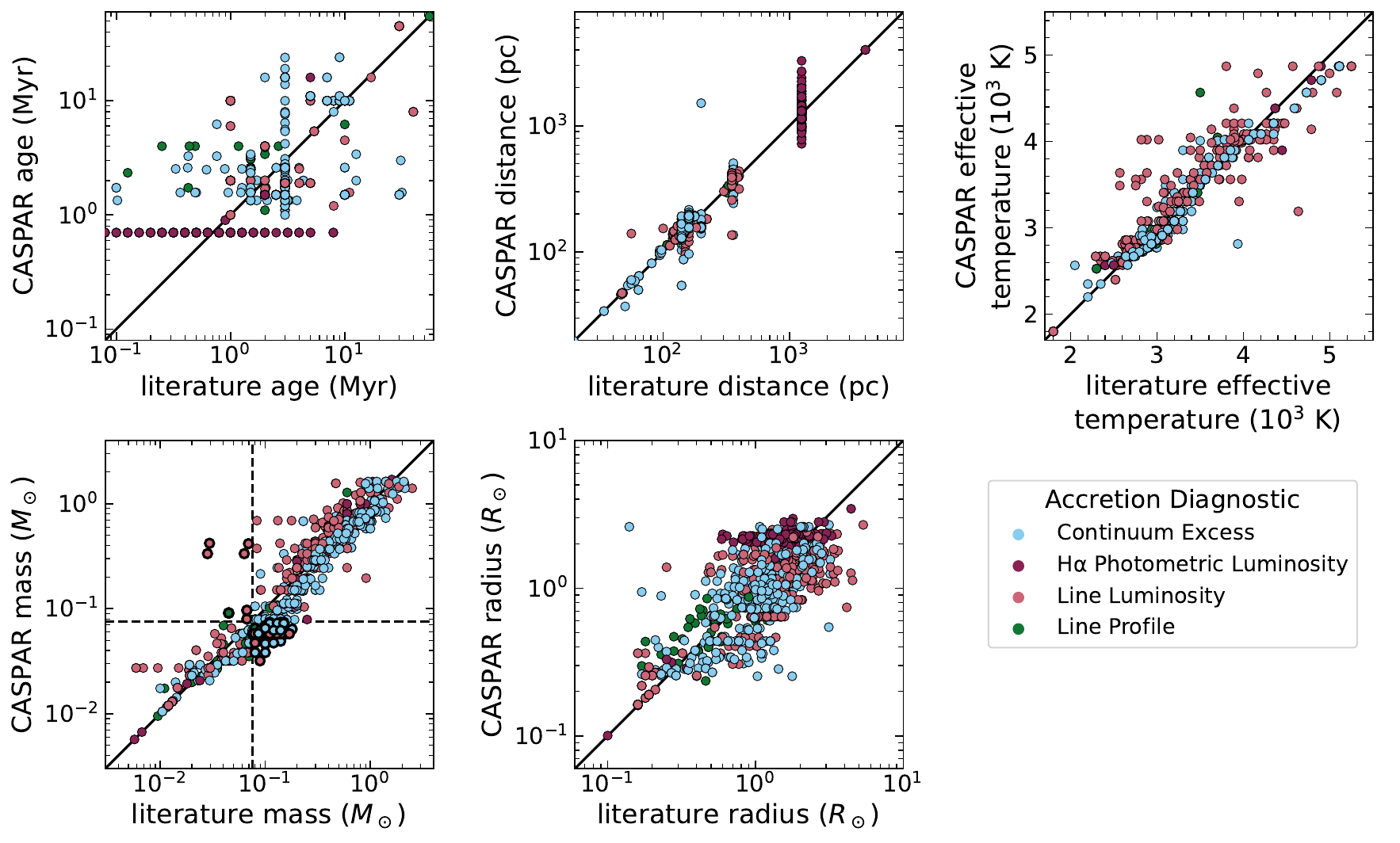}
    \caption{Comparison between the Literature Database physical parameters and those rederived in CASPAR colored by accretion diagnostic.  The black line indicates 1-1.  For masses, we show the hydrogen burning limit at $M = 0.075\ M_\odot$ with dashed black lines; under a uniform methodology, objects in the lower right quadrant decreased in mass from stellar to substellar, while in the upper left, the reverse occurred.  
    We find $50-89$\% of CASPAR parameters are within a factor of two of the literature value.}
    \label{fig3:refCASPAR_physicalcomparison}
\end{figure*}

We show comparisons of rederived CASPAR parameters and the literature parameters in Fig.~\ref{fig3:refCASPAR_physicalcomparison}.  For all rederived quantities, 50--89\% are within a factor of two of the literature value, indicative of the relatively small effects of these updates.  The change from individual or previous estimates of star forming region ages to a uniform set of ages result in the large dispersion seen between the original and rederived values. 

We find 60 objects that were originally considered low mass stars with masses $>0.075\ M_\odot$ are now classified as brown dwarfs with masses below the hydrogen burning limit (HBL; conversely six of the brown dwarfs are now classified as stars). We highlight these in the lower right quadrant in Fig.~\ref{fig3:refCASPAR_physicalcomparison}. Of those, $\sim20$ are within 1$\sigma$ of the HBL.  These mass shifts result from using the \citet{Herczeg2014} spectral type to temperature conversion and updated ages that were not used by the original references. Of the 16 studies where objects traverse below the HBL, 11 were published before the \citet{Herczeg2014} spectral type-temperature conversion; therefore, their methods of calculating temperature and spectral types differed (with the majority using the conversion of \citet{Luhman2003}) leading to differences when reevaluating them.  The four published after \citet{Herczeg2014} utilized their own spectral fitting and used the spectral type-temperature conversion of \citet{Luhman2003} for the M dwarfs in their sample.  When we look at the temperature -spectral type conversions from \citet{Herczeg2014} and \citet{Luhman2003}, we find that they diverge $\sim100-150$ K for M dwarfs, a change of $\sim0.06\ M_\odot$, with the \citet{Herczeg2014} temperatures found to be lower for the same spectral type. 
This results in the decrease in mass seen in CASPAR for these objects.

The objects are from 12 different star forming regions and from 16 different original references, indicating no preferential biases in deriving masses.  Under this uniform derivation, we find 658 stars, 130 brown dwarfs, and 10 PMCs (see Table~\ref{tab:numbercounts}).

\begin{deluxetable}{lcc}
\tablecaption{Population counts between literature database and CASPAR\label{tab:numbercounts}}
\tablewidth{0pt}
\tablehead{\colhead{Population} & \colhead{Literature Database} & \colhead{CASPAR}}
\startdata
Star & 712 & 658 \\
BD & 76 & 130 \\
PMC & 10 & 10 \\
\enddata
\end{deluxetable}

\begin{figure}[thp]
    \centering
    \includegraphics[width=.96\linewidth]{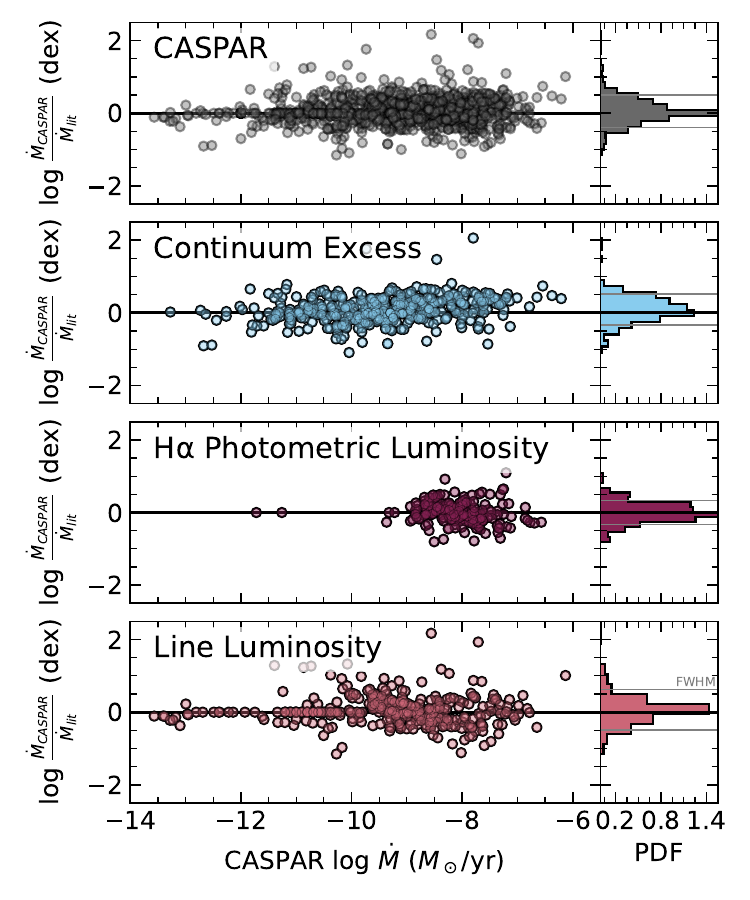}
    \caption{Residuals between reference and CASPAR accretion rates as a function of CASPAR accretion rate for different accretion diagnostics.  88\% of all rederived values changed by less than 0.5 dex, with the line luminosity accretion diagnostic showing the most change in calculated accretion rate. \textit{right:} Histogram of difference between literature and rederived accretion rates for each tracer. The thin gray lines indicate the FWHM.  Colors as in Fig.~\ref{fig3:refCASPAR_physicalcomparison}.}
    \label{refCASPAR_accretion}
\end{figure}

The residuals between CASPAR accretion rates and their literature-derived values are shown in Fig.~\ref{refCASPAR_accretion} for the full sample and each accretion diagnostic.  Overall, we find that 1$\sigma$ standard deviation in the residuals for all rederived accretion rates is 0.38 dex over the range of [$-3.8$, 1.15] dex (mean $= -0.05$ dex). We find that 909 (88\%) of the accretion rates change by less than 0.5 dex, indicating the vast majority of the objects have not markedly changed from their literature value. We find that the most disparate changes in accretion rate measurements are by H$\alpha$ photometric luminosity and continuum excess, with CASPAR measurements larger than previously calculated.  
When we look at those accretion rates calculated from excess continuum, we find a median difference between the rederived and literature radius/mass ratio of 0.3 dex.  As these accretion rate measurements are dependent on $R/M$ ($\dot M \propto R/M$), this in turn leads to a median difference between the rederived and literature $\dot M$ of $\sim0.1$ dex, indicating the majority of accretion rates derived from excess continuum are higher than originally estimated.

\section{Linear Fitting Technique}\label{sec:linfit}
With CASPAR, we can better investigate the causes of scatter in accretion rates. In the following sections, we discuss the relationship between accretion rate and parameters such as mass and age. Unless otherwise stated, these relationships are derived using the hierachical Bayesian linear regression routine \texttt{linmix} \citep{Kelly2007} in Python\footnote{https://github.com/jmeyers314/linmix} to determine the slope, intercept, and intrinsic scatter around relations of the form: $y = mx + b$.  \texttt{linmix} allows for heteroscedastic and correlated measurement errors, and takes upper limits into account.  It assumes that $x$ and $y$ variables are drawn from a 2D Gaussian distribution, and the covariance matrix is composed of the uncertainties in $x$ and $y$.  Calculated regression coefficients and uncertainties are derived from the posterior probability distributions of model parameters computed using Markov Chain Monte Carlo (MCMC).  

Not every object in the Literature Database, and therefore CASPAR, has a reported $\dot M$ uncertainty.  For objects with no literature uncertainty, we assume the average $\dot M$ uncertainty from all reported and rederived measurements in CASPAR ($\sim$ 0.36 dex) in the fit.      
  
For each fit performed, we recover the posterior distribution of the slopes and intercepts and calculate best fit parameters from the median and $1\sigma$ uncertainties.  Additionally, we calculate the Pearson correlation coefficient, $R$, between the $x$ and $y$ parameters. Best fit coefficients are recorded in Table~\ref{tab:best fit}.

\begin{deluxetable*}{l|lccccc}
\tablecaption{Best fit parameters\label{tab:best fit}}
\tablewidth{0pt}
\tablehead{\colhead{} & \colhead{} & \colhead{N$^*$} & \colhead{$a$ ($\pm$err)} & \colhead{$b$ ($\pm$err)} & \colhead{$\sigma^\dag$} & \colhead{R}}
\startdata
\multicolumn{7}{c}{Literature $\dot M \sim M$ fits}		\\
\hline
\hline
&Total	&	1038	&	2.15	(0.08)	&	$-$8.03	(0.06)	&	1.01	&	0.73	\\
\hline
By Mass &Star	&	837	&	2.69	(0.13)	&	$-$7.80	(0.07)	&	0.92	&	0.64	\\
\hline
& Continuum Excess & 396 & 1.92(0.13) & $-$8.53(0.09) & 0.87 & 0.66 \\ 
By Accretion & H$\alpha$ Photometric Luminosity & 223 & 1.54(0.19)&$-$7.47(0.08) & 0.05 & 0.84 \\
Diagnostic & Line Luminosity & 346 & 1.99(0.11) & $-$8.24(0.12) & 1.10 & 0.74 \\
 & Line Profile & 73 & 0.96(0.46) & $-$9.87(0.60) & 2.00 & 0.27 \\
\hline	
\multicolumn{7}{c}{CASPAR $\dot M \sim M$ fits}	\\
\hline
\hline
&Total	&	1038	&	2.02	(0.06)	&	$-$8.02	(0.05)	&	0.85	&	0.76	\\
\hline
& Star	&	764	&	2.17	(0.11)	&	$-$7.97	(0.05)	&	0.74	&	0.63	\\
& Star + BD	&	1000	&	2.12	(0.07)	&	$-$7.98	(0.05)	&	0.85	&	0.76	\\
By Mass & BD	&	236	&	3.19	(0.57)	&	$-$6.54	(0.80)	&	1.36	&	0.40	\\
& BD + Planet	&	275	&	1.55	(0.37)	&	$-$8.75	(0.55)	&	1.36	&	0.29	\\
& Planet	&	38	&	0.25	(2.91)	&	$-$10.66	(5.64)	&	0.81	&	0.04	\\
\hline
& Continuum Excess & 396 & 1.87(0.11) & $-$8.31(0.09) & 0.92 & 0.68 \\ 
& H$\alpha$ Photometric Luminosity & 223 & 1.25(0.18)&$-$7.69(0.06) & 0.13 & 0.64 \\
& Line Luminosity & 346 & 1.76(0.10) & $-$8.34(0.11) & 1.15 & 0.71 \\
By Accretion & Line Profile & 73 & 1.64(0.40) & $-$8.91(0.52) & 1.56 & 0.46 \\
Diagnostic & Balmer	&	1558	&	2.18	(0.04)	&	$-$7.81	(0.04)	&	0.51	&	0.85	\\
& Pashen/Brackett/Pfund	&	644	&	1.40	(0.06)	&	$-$8.17	(0.05)	&	0.37	&	0.77	\\
& HeI	&	486	&	1.72	(0.07)	&	$-$8.09	(0.07)	&	0.49	&	0.76	\\
& CaII	&	412	&	1.74	(0.08)	&	$-$8.03	(0.08)	&	0.41	&	0.80	\\
\hline
& Total $\leq$1 Myr	&	231	&	2.13	(0.17)	&	$-$7.45	(0.07)	&	0.13	&	0.85	\\
& Total (1$-$3] Myr	&	578	&	1.78	(0.08)	&	$-$8.23	(0.07)	&	0.91	&	0.69	\\
& Total (3$-$8] Myr	&	59	&	1.51	(0.21)	&	$-$8.51	(0.23)	&	0.78	&	0.73	\\
& Total $>$8 Myr	&	169	&	1.64	(0.15)	&	$-$8.95	(0.17)	&	1.20	&	0.67	\\
By Age & Star $\leq$1 Myr	&	224	&	1.53	(0.40)	&	$-$7.61	(0.12)	&	0.14	&	0.45	\\
and Mass & Star (1$-$3] Myr	&	413	&	2.12	(0.14)	&	$-$8.11	(0.18)	&	0.76	&	0.64	\\
 & Star (3$-$8] Myr	&	29	&	2.43	(0.66)	&	$-$8.21	(0.32)	&	1.09	&	0.61	\\
& Star $>$8 Myr	&	97	&	1.21	(0.31)	&	$-$9.12	(0.21)	&	1.36	&	0.37	\\
& BD $\leq3$ Myr	&	158	&	1.68	(0.79)	&	$-$8.41	(1.10)	&	1.57	&	0.2	\\
& BD (3$-$8] Myr	&	21	&	5.88	(4.64)	&	$-$3.06	(5.92)	&	0.77	&	0.46	\\
& BD $>$8 Myr	&	57	&	4.44	(0.76)	&	$-$5.20	(1.12)	&	0.64	&	0.68	\\
\hline
\multicolumn{7}{c}{$\dot M \sim$ Age}	\\
\hline
\hline
\multicolumn{2}{c}{Literature}	&	1038	&	$-$0.84	(0.06)	&	$-$3.17	(0.38)	&	0.80	&	$-$0.46	\\
\multicolumn{2}{c}{CASPAR} 	&	1038	&	$-$0.86	(0.09)	&	$-$2.81	(0.60)	&	0.77	&	$-$0.31	\\
\enddata
\tablecomments{Linear fit in the form: $\log \dot{M}=a\times X + b$, where $X$ is either $\log M$ ($M_\odot$) or Age (Myr).}
\tablenotetext{*}{N refers to the number of $\dot M$ measurements}
\tablenotetext{$\dag$}{standard deviation of the linear fit.}
\end{deluxetable*}

In Appendix~\ref{app:fittingtechniques}, we discuss additional linear regression methods that were investigated, such as weighted least squares, ordinary least squares bisector, and orthogonal distance regression.  We find that including upper limits into the fits does not significantly affect the best-fit coefficients.  However, the inclusion of $x$ axis uncertainties can significantly affect the resulting best-fit coefficients. 

For fits that do not follow $y=mx+b$ (such as $y=e^x$) and therefore cannot be fit with \texttt{linmix}, we utilize the orthogonal distance regression code \texttt{scipy.odr}, which allows for fitting of non-linear functional forms while taking into account both $x$ and $y$ measurement uncertainties, but not upper limits (though this should not greatly affect the fit as discussed above).

\section{Effect of methodology on Accretion Rate Scatter}\label{sec:scatter}
CASPAR spans a wide range of masses from $\sim$10 $M_\mathrm{J}$ to $\sim2\ M_\odot$ and compiles accretion rates calculated from four diagnostics.  We first investigate a) the extent to which a uniform derivation reduces the extensive scatter in $\dot M$, and b) whether or not the accretion diagnostic influences this scatter. 

\subsection{Scatter in Accretion Rate after Re-derivation}\label{scatter1}
The relationship between mass accretion rate and mass follows a power law of the form $\dot{M}\propto M^{\alpha}$, where previous measurements of $\alpha$ range from 1.0--2.8 \citep[e.g.][]{Calvet2004, Natta2004, Mohanty2005, Muzerolle2005, Herczeg2008, Zhou2014, Hartmann2016}. Typical dispersions are 1--2 dex \citep[][and references therein]{Manara2022}. The top panel of Fig.~\ref{fig:origIndiv_lines_mmdot} shows the best fit to the Literature Database as $\log\dot{M}=2.16(\pm0.08)\log M-8.03(\pm0.06)$, with a $1\sigma$ dispersion of 1.00 dex and correlation of 0.73. The wide range of previously measured stellar slopes is consistent with our Literature Database fit, and is likely driven by uncertainties in the masses and differences in sample ages and sizes \citep{Hartmann2016}. 

\begin{figure*}[tb]
    \centering
    \includegraphics[width=\linewidth]{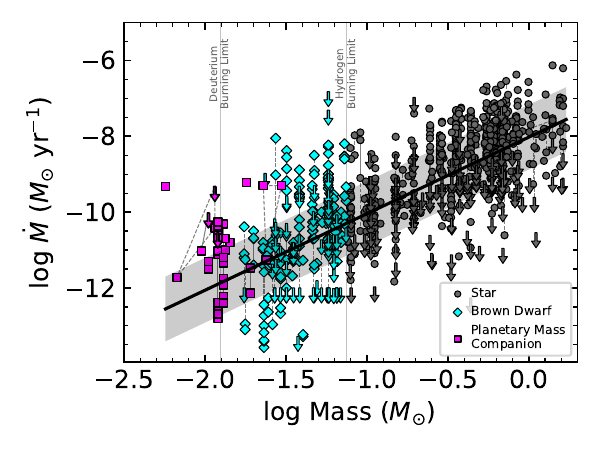}
    \caption{The CASPAR $\dot M-M$ relation for stars (black circles), brown dwarfs (cyan diamonds), and PMCs (magenta squares).  
    The gray dashed lines show accretion rates derived for the same object, and the downward arrows show accretion rate upper limits. The black line and shaded region shows our best linear fit, $\log \dot M = 2.02 \log M - 8.02$, and 1$\sigma$ dispersion, $\sigma=0.85$~dex ($\dot M$ in $M_\odot\,\textrm{yr}^{-1}$ and $M$ in $M_\odot$), to all accretion rates in CASPAR.} 
    \label{fig:bestfits}
\end{figure*}

Therefore, we first 
affirm that the rederived values in CASPAR are consistent with previous slope estimates and discuss how the scatter changes for a larger sample and under a uniform methodology. We show the best fit model for CASPAR $\dot{M}-M$ in Fig.~\ref{fig:bestfits} and find a linear trend of:
\begin{equation}
    \log \dot{M} = 2.02(\pm0.06)\log M - 8.02(\pm0.05),
\end{equation}
with a 1$\sigma$ dispersion of 0.85 dex and a strong correlation coefficient of 0.76.

CASPAR was based on previous studies; therefore, a consistent slope with literature estimates is unsurprising.  A uniform derivation reduces the $1\sigma$ dispersion by 0.15 dex, indicating that methodology accounts for only 17\% of the original scatter.  
This implies that the remaining scatter is due to underlying physical mechanisms rather than methodological systematics, and that methodology (e.g., object mass and radius estimation technique, varied distance references, etc.), while accounting for some of the dispersion in the scatter, cannot explain all of it, especially as large scatter is seen in uniform star forming region surveys \citep[e.g.][]{Manara2015, Alcala2017}.

\begin{figure*}[htp!]
    \centering
    \includegraphics[width=\linewidth]{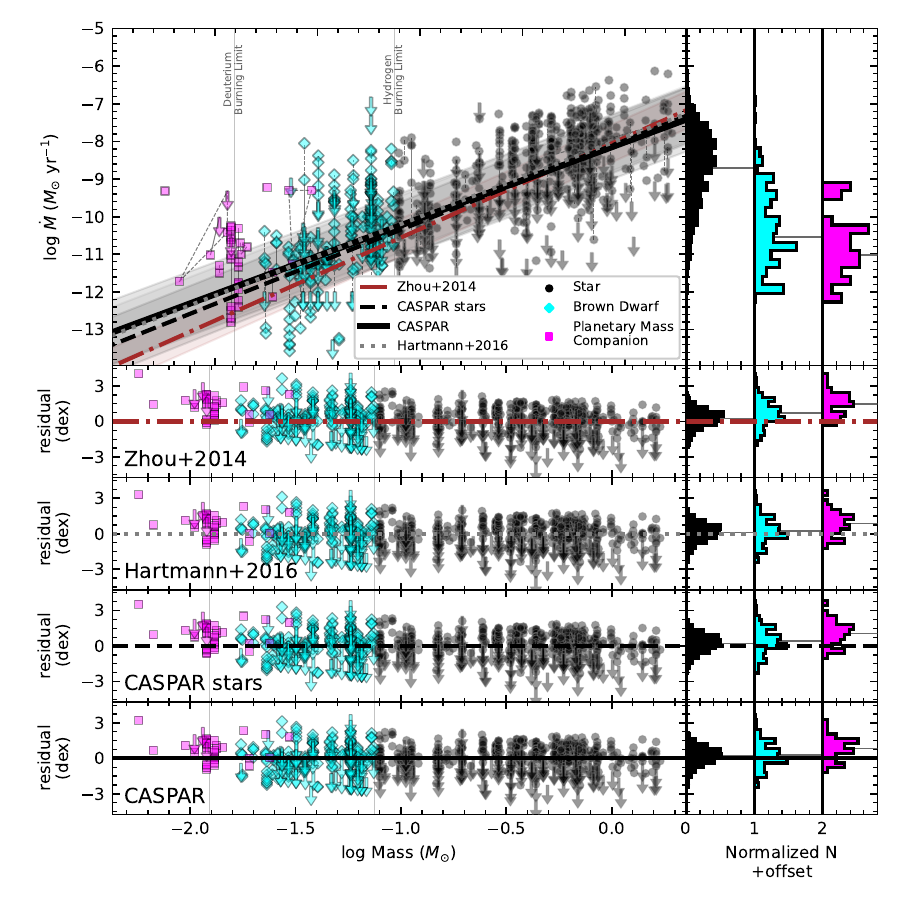}
    \caption{\textit{left}: The CASPAR $\dot M-M$ best-fit relation for overall CASPAR (black solid line), CASPAR stars-only (black dashed dot line), \citet{Zhou2014} (gray dotted line), and \citet{Hartmann2016} (brown solid line). The 1$\sigma$ scatter is shown as shaded regions for each fit. 
    The residuals between each linear fit and CASPAR are shown below.  \textit{right}: Histogram of residuals for each population.  The thin gray lines shows the mean of each distribution. Markers and colors are as in Fig.~\ref{fig:bestfits}.}
    \label{fig:new_MMdotResiduals}
\end{figure*}

In Fig.~\ref{fig:new_MMdotResiduals}, we also compare the best fit $\dot M-M$ relations from \citet{Zhou2014} and \citet{Hartmann2016} to CASPAR. Though we do not expect significant variation between our fit and previous estimates, the samples from \citet{Zhou2014} and \citet{Hartmann2016} were primarily composed of stars between 0.1 and 1 $M_\odot$, while our fit includes objects down to $0.01\ M_\odot$.  Therefore, we investigate whether the substellar population is different than previous stellar fits. 

We compare the best fit relation of (a) the CASPAR total sample and (b) stellar sample to the fits of \citet{Hartmann2016} and \citet{Zhou2014}.  We find that the \citet{Zhou2014} and CASPAR star-only ($0.075 < M / M_\odot < 1.7$) fits show positively skewed residuals when extended into the brown dwarfs and PMC mass regimes, while the \citet{Hartmann2016} and CASPAR single population fit shows a positive skew only for PMCs (due to the stellar population dominating the fit).  
In the histograms of the residuals in Fig.~\ref{fig:new_MMdotResiduals}, we indicate the center of the distribution with a narrow gray line. We find that while the four fits overlaps in the stellar regime, they deviate at substellar masses, with our fits accounting for most of the scatter, within fit uncertainties.

\subsection{Accretion Diagnostic Systematics} \label{acc diag systematics}

\begin{figure*}[tb!]
    \centering
    \includegraphics[width=\linewidth]{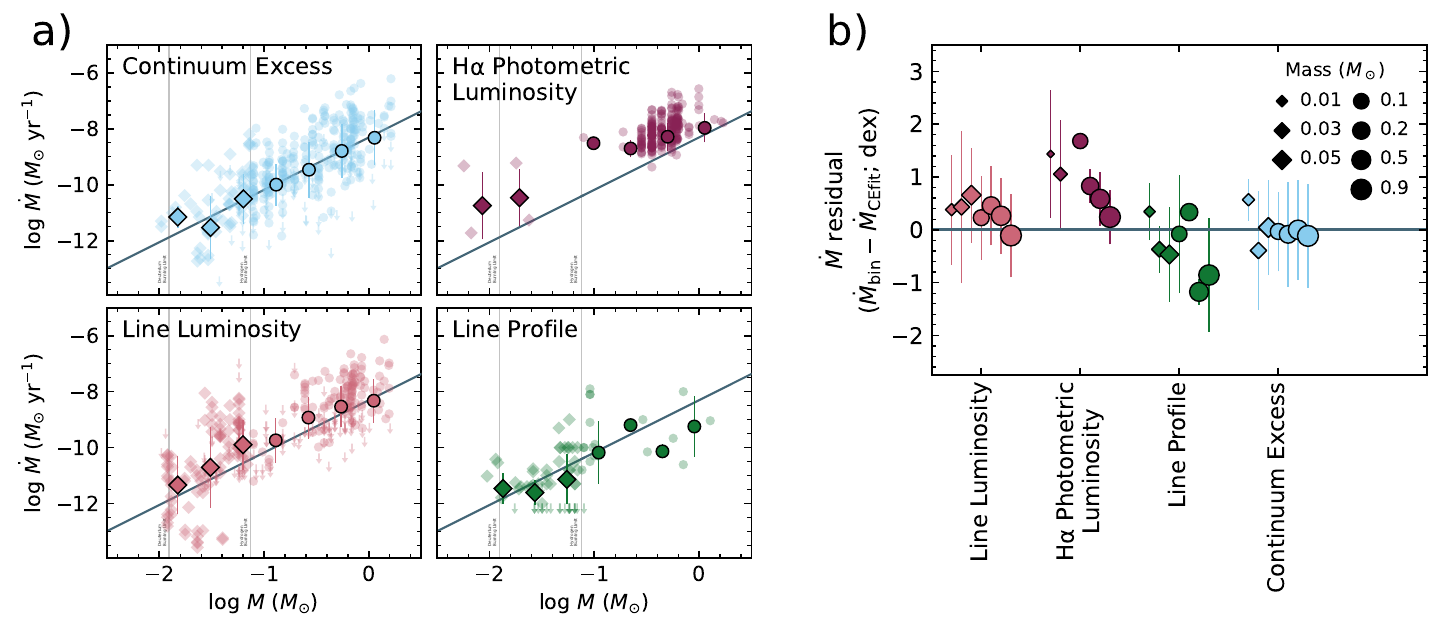}
    \caption{\textit{Panel a.} Accretion rate vs mass for individual accretion diagnostics colored as in Fig.~\ref{fig:origIndiv_lines_mmdot}. The large markers indicated the binned accretion rates by mass. \textit{Panel b.} Residuals between binned accretion rates and CASPAR continuum excess fit.  Line luminosities (including H$\alpha$ photometric) show (significant) deviation from the continuum excess with decreasing mass. H$\alpha$ photometric luminosities appears significantly offset from the CASPAR continuum excess best fit; however, the majority of the stellar accretion rates are from objects with ages $<1$ Myr, which biases the high mass results (see Sec~\ref{AllAccretionAge}). }
    \label{fig:binbyad}
\end{figure*}

To assess the consistency among various observational methods, we assume that all methods should produce consistent $\dot M$ values of estimating accretion rates. In Fig.~\ref{fig:binbyad}a, we show CASPAR $\dot M-M$ statistics separated by accretion diagnostic.  We bin accretion rates by mass to facilitate comparison.  To do this, we assume every detection is a Gaussian probability density distribution (PDF) with a mean of the accretion rate and standard deviation of the uncertainty. For non-detections, we assume a half normal distribution with the cut-off at the upper limit value.  We then ran a Monte Carlo simulation drawing random values from the PDFs within each mass bin and took the median of the values.
We find consistent median accretion rates for each mass bin across the four diagnostics, with the medians closely following the best fit line found for continuum excess. Fig.~\ref{fig:binbyad}b shows the fit residuals. Though within uncertainties, H$\alpha$ photometric luminosities produce higher ($\sim$1 dex) $\dot M$s, while line profiles produce lower ($\sim$1 dex) ones. However, the majority of the H$\alpha$ photometric measurements are for objects with ages $<1$ Myr, biasing the results (see Section~\ref{AllAccretionAge}). Overall, line luminosity and continuum excess methods produce the smallest residuals in the stellar regime (residuals $< 0.4$ dex) of the four diagnostics.  As there are only 10 PMCs, in which the masses and accretion rates are not uniformly derived, we cannot conduct a similar analysis, though we show them in Fig.~\ref{fig:binbyad}.  

We find line profile derived accretion rates lie consistently below the best fit line for the substellar mass objects, while continuum excess is consistent to within $\sim0.21$ dex across all masses.  For line luminosities, we find relatively small residuals (up to 0.8 dex) in the substellar regime.  Though within uncertainties, line luminosity and line profile residuals trend upward from the stellar to substellar regime, while this is not seen for the accretion rates derived from the continuum excess tracers.  The standard deviation of the residuals between line luminosity and excess continuum derived rates increase from 0.1 dex at $0.5\ M_\odot$ to 0.65 dex at $0.03\ M_\odot$.  
As line luminosities are dependent on scaling relations (derived from excess continuum) to compute accretion rates, we expect similar trends among excess continuum and line luminosity derived $\dot M$s.  While they are consistent within uncertainties, the average accretion rates do show differences at low masses, potentially a result of utilizing stellar scaling relations in the substellar regime. We discuss this hypothesis further in Section~\ref{sec:BDpop}.       

Accretion rates measured from line emission are known to have significant uncertainty, as scaling relations are empirically derived and many accretion tracing lines can also be produced by other physical processes, such as winds and chromospheric activity \citep{Jayawardhana2003, White2003}. Several criteria have been established to try to separate accretion and chromospheric activity using H$\alpha$ equivalent widths \citep[$<200$ km/s;][]{Jayawardhana2003, White2003} and $L_\mathrm{acc}/L$ ratio as a function of temperature (and spectral type) for emission lines \citep[$-3.19\pm0.15$ for M6 dwarfs;][]{Manara2017}.

\begin{figure*}[htp!]
    \centering
    \includegraphics[width=.95\linewidth]{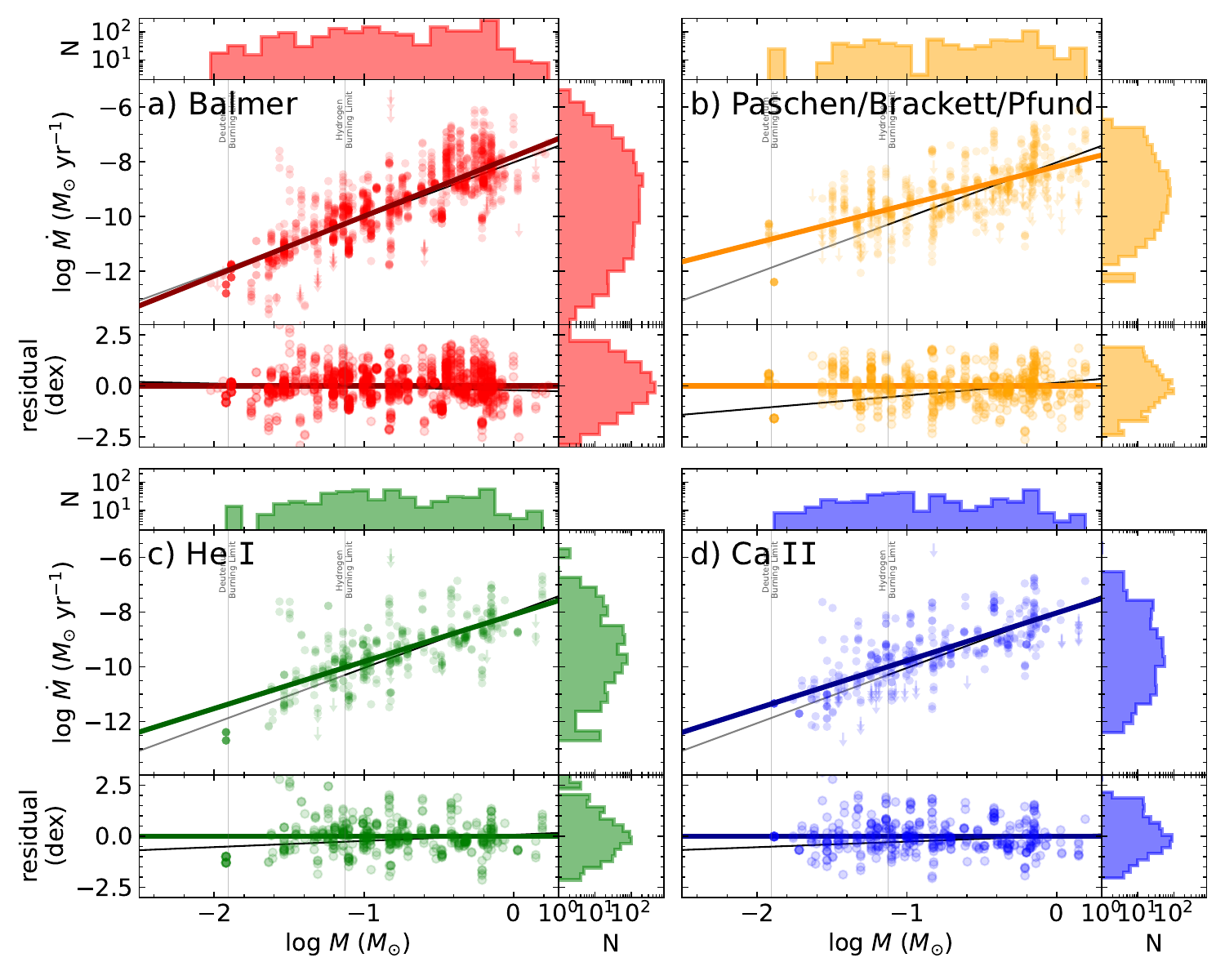}
    \caption{Accretion rate vs mass for a range of accretion-tracing lines, namely: a) optical Hydrogen Balmer series lines b) NIR Hydrogen Paschen, Brackett and Pfund series lines, c) Helium~\textsc{i} emission lines and d) Calcium~\textsc{ii} emission lines. The gray line shows the best linear $\dot M-M$ fit to the overall CASPAR database, while the colored lines show the best linear fit for each emission line. The histograms show the posterior distributions for the masses (top), accretion rates (top right), and accretion rate residuals (bottom right) for each panel. We find that the NIR line flux fit shows the most deviation in best fit; it is offset by 0.9 dex compared to the best linear fit at the deuterium burning limit (compared to $<0.4$ in the other lines).}
    \label{fig:byline}
\end{figure*}

In Fig.~\ref{fig:byline}, we group line fluxes by wavelength, namely: a) Balmer series, b) infrared hydrogen series (Paschen, Brackett and Pfund), c) Helium~\textsc{i} emission lines (He~\textsc{i} $\lambda$\,4026, He~\textsc{i} $\lambda$\,4471, He~\textsc{i} $\lambda$\,4713, He~\textsc{i} $\lambda$\,5016, He~\textsc{i} $\lambda$\,5876, He~\textsc{i} $\lambda$\,6678, He~\textsc{i} $\lambda$\,7065), and d) Calcium~\textsc{ii} emission lines (Ca~\textsc{ii} K, Ca~\textsc{ii} H, Ca~\textsc{ii} $\lambda$\,8498, Ca~\textsc{ii} $\lambda$\,8542, Ca~\textsc{ii} $\lambda$\,8662), in order to analyze trends among them.
We first find best fit $\dot M-M$ relations for each line flux group (given in Table~\ref{tab:best fit}) and compare these fits to the CASPAR best fit relation. 
We use the Akaike’s information criterion (AIC) to access the performance of the linear fits in explaining the variation in the data\footnote{The AIC, defined as AIC~$=2k - 2\mathrm{ln}(L)$, informs the relative quality of different models against a given set of data, where $k$ is the number of estimated model parameters, and $L$ is the likelihood function of the model.} Since our two fits are independent (i.e., non-nested) with the same number of parameters, this criterion estimates how well the model reproduces the data from the maximum likelihood.

If the AIC values for two fits are within 10\%, we consider the fits to be comparable. If the fit to a line flux group is not significantly more descriptive of the variance than the overall best fit relation, we conclude that offsets by method do not contribute to the scatter. 
In the stellar regime, we find the AICs for each line flux group are all within 10\%.  

In the substellar regime, the best fit line to the infrared hydrogen and He~\textsc{i} line measurements is offset from the overall best fit with a percent difference between AIC $\sim180$\%.  Additionally, at the deuterium burning limit, the NIR best fit line is offset from the CASPAR best fit by 0.9 dex (compared to $<0.4$ dex for other lines).  This could indicate that the NIR line flux scaling relations overestimate accretion rates in the substellar regime (i.e. their scaling relations are different), or this may result from small number statistics.

Finally, we estimate the extent to which accretion rates derived from H$\alpha$ 10\% widths, H$\alpha$ line profiles, and UV excess differ in order to probe their effect on scatter. When accretion rates are derived from different emission lines or continuum at the same epoch, they should not be subject to intrinsic accretion variability. This makes contemporaneous measurements an excellent probe of systematics.  While $\dot M$s from H$\alpha$ luminosity rely on $L_\mathrm{line}-L_\mathrm{acc}$ scaling relations, the $\dot M-$H$\alpha$ 10\% width scaling relation of \citet{Natta2004} relates the 10\% width directly to the accretion rate.  However, \citet{Alcala2014} found that it can underestimate accretion rates by almost 0.6 dex for widths $<400$ km/s in Lupus (corresponding to $M<0.3\ M_\odot$) compared to excess continuum.  We find a similar result when we compare the CASPAR H$\alpha$ 10\% widths to accretion rates derived from excess continuum.  

In Fig.~\ref{fig:Ha10HaUV}, we show the residuals in CASPAR accretion rates derived from excess continuum (top) and H$\alpha$ 10\% width (bottom) compared to simultaneous measurements from H$\alpha$ luminosity as a function of mass. In the stellar regime, accretion rates derived from these three quantities do not significantly differ within 1 dex. 
In the substellar regime, accretion rates derived from H$\alpha$ luminosity are systematically high when compared to excess continuum, which could be indicative of an overestimation of accretion luminosity for the lowest mass BDs and PMCs. 

We find a large offset in accretion rates derived from H$\alpha$ 10\% width compared to H$\alpha$ luminosity. Several other processes, including chromospheric activity, outflows, and hotspots, contribute to H$\alpha$ emission, potentially inflating its width and therefore increasing its inferred accretion rate. Low mass substellar accreting objects can have line widths below the traditional 200 km/s threshold for accretion.  Such observations have led previous work \citep[e.g.][]{Alcala2014, Alcala2017} to discourage the use of H$\alpha$ 10\% width as an accretion tracer.  We confirm this offset between H$\alpha$ 10\% width and H$\alpha$ luminosity within the substellar regime and find it can lead to a difference of almost 2 dex in calculated accretion rate near the deuterium burning limit, producing vastly overestimated accretion rates.

\subsection{Summary}
We find that methodological differences in estimates of mass accretion rates, such as differences in evolutionary models and estimated distances (used to scale accretion luminosities), account for only $\sim17$\% of the scatter in the $\dot M-M$ relation, indicating that the remaining scatter is from either observed accretion diagnostic systematics or physical differences (e.g., variability, disk mass, stellar mass, system age). 

\begin{figure}[htp!]
    \centering
    \includegraphics[width=\linewidth]{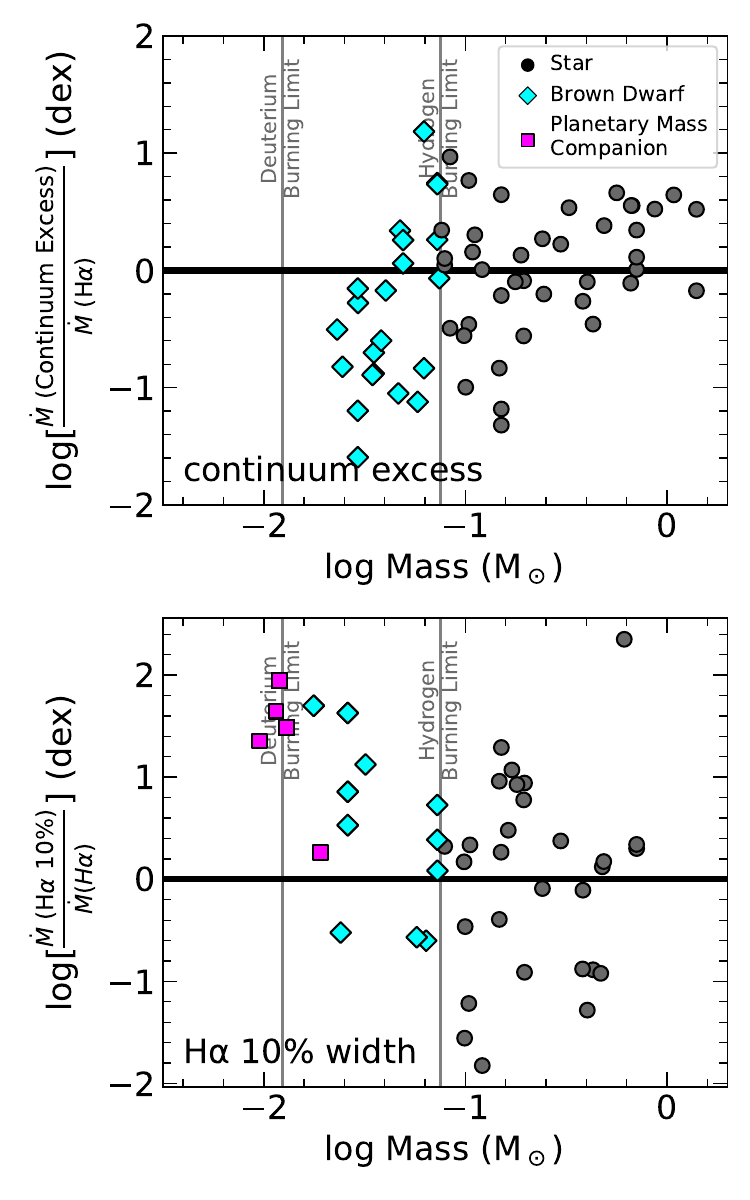}
    \caption{Difference in accretion rates derived from excess continuum (top) and H$\alpha$ 10\% width (bottom)  compared to H$\alpha$ luminosity as a function of mass for objects with simultaneous H$\alpha$ luminosity and either excess continuum or H$\alpha$ 10\% width. As mass decreases, we find an offset in accretion rate between these quantities.  Accretion rates derived from these simultaneous observations should not physically vary, thereby underscoring systematic methodological differences in calculating accretion rates.   }
    \label{fig:Ha10HaUV}
\end{figure}

When we separate accretion rate estimates by accretion diagnostic, we find that accretion rates do not vary significantly ($<1$ dex) in the stellar or substellar regime.  However, we do find systematically higher accretion rates for $\dot M$s derived from NIR line luminosities (0.9 dex offset in best fit line at the deuterium burning limit; Fig.~\ref{fig:byline}b) and H$\alpha$ luminosity relative to continuum estimates (top panel of Fig.~\ref{fig:Ha10HaUV}).  We will discuss the physical and diagnostic drivers of this scatter in the following sections.  

\section{Drivers of Physical Scatter in Mass Accretion}\label{sec:physicalscatter}
As shown in the previous section, methodological systematics cannot fully explain the scatter in the $\dot M-M$ relation for either stellar or substellar objects. Recent work has suggested that multiplicity may affect accretion rates, with binaries accreting at a higher rate than isolated objects \citep{Zagaria2022, Gangi2022}.  While CASPAR currently does not contain many objects in multiple systems (46/798), they appear consistent with isolated object accretion rates and do not have a significant effect on the $\dot M-M$ scatter.  Below, we focus on the relationship between age and intrinsic variability in the observed $\dot M-M$ scatter\footnote{Though disk mass has a known correlation with accretion rate \citep[][and references therein]{Manara2022}, and modeling work has suggested variations in accretion rates are due to differences in disk mass \citep{Vorobyov2009}, disk masses are not currently collected in CASPAR. We cross-match CASPAR with the sample from \citet{Manara2022} and reproduce the \citet{Manara2022} results comparing their disk masses with CASPAR stellar masses, ages, and accretion rates. Therefore, we focus here on quantifying age and multi-epoch variability.}, and what this may tell us about the evolution of accretion activity.  

\subsection{Variation in Accretion with Age}\label{AllAccretionAge}

Circumstellar disk fraction in young star forming regions has been found to decrease exponentially with age until $\sim$10-15~Myr \citep{Mamajek2009, Luhman2022}, with the majority of disks dissipating after $\sim2.5$~Myr. The rate of decay is mass dependent, as \citet{Luhman2022} found an increase in disk fraction with decreasing mass in the $15-21$~Myr Lower Centaurus Crux and Upper Centaurus Lupus associations.  They found that the lower mass objects retained their disks longer than the previously presumed disk dispersal timescales of 10--15~Myr.  

Correlations among accretion rates, disk gas masses, and accretion timescales \citep{Hartmann1998, Manara2016b, Mulders2017} are generally explained through a combination of viscous evolution \citep{Lodato2017, Rosotti2017, Mulders2017}, disk photoevaporation \citep{Sellek2020}, and stellar multiplicity \citep{Zagaria2022}.            
Accretion rates decrease with time as $t^\alpha$, where $\alpha=-1.6$ to $-1.2$ \citep[][and references therein]{Hartmann2016}.  This decrease has been attributed to viscous evolution, though observations of the POISSON sample found higher $\dot M$ than expected from pure viscous models \citep{Antoniucci2014}.  

By establishing a ``uniform" estimate of ages, we can study the correlation of $\dot M$ with age, and its impact on $\dot M-M$ scatter. We exclude PMCs in this analysis as a) they have not been uniformly re-derived, and b) theoretically modeled accretion \citep{Aoyama2018, Thanathibodee2019, Aoyama2020} and formation \citep{Stamatellos2015} mechanisms posit that stellar age trends could not hold for PMCs. We divide CASPAR into the following four age bins for this analysis:  
\begin{itemize}
    \item $\leq1$ Myr: includes the Lagoon Nebula
    \item $1 < t/\mathrm{Myr} \leq 3$: includes Chamaeleon I, Taurus, and Lupus
    \item $3 < t/\mathrm{Myr} \leq 8$:  includes IC 348 and Perseus
    \item  $> 8$ Myr: includes Upper Centaurus Lupus and $\eta$ Chamaeleontis.
\end{itemize} 
See Table~\ref{tab:ages} for full list of ages for each cluster and association within these broad groups.   

\begin{figure}[htp!]
    \centering
    \includegraphics[width=\linewidth]{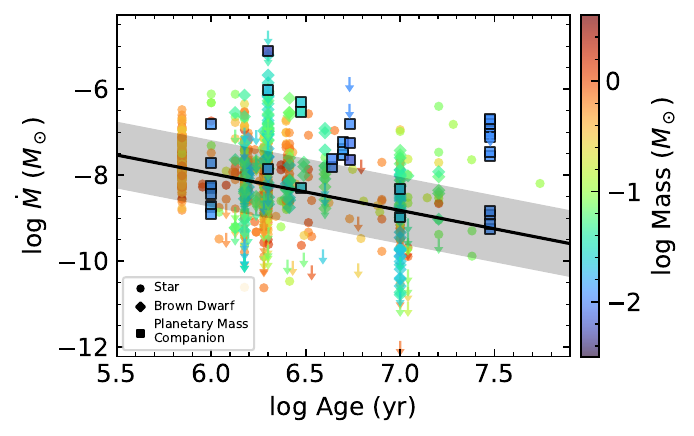}
    \caption{CASPAR accretion rate vs age colored by mass. The accretion rates have been scaled by the $\dot M-M$ relation to $M\sim 0.7\ M_\odot$ following \citet{Hartmann2016}.  The black line and shaded region show the best linear fit and 1$\sigma$ scatter.  }
    \label{fig:MdotAge}
\end{figure}

In Fig.~\ref{fig:MdotAge}, we show accretion rate as a function of age, colored by stellar mass. Following \citet{Hartmann2016}, we scale to the accretion rate to $M_*=0.7\ M_\odot$ in order to remove the dependence on mass. The best fit to these data using \texttt{linmix} is:
\begin{equation}
    \log \frac{\dot M}{M_\odot/\mathrm{yr}} = -0.85(\pm 0.09)-2.80(\pm0.60)\times \log \frac{t}{\mathrm{yr}},
\end{equation}
with a scatter of 0.77 dex and Pearson correlation coefficient of $-0.31$. 
The slope found by \citet{Hartmann2016} (slope $=-1.07$) shows a faster decline in accretion rate with age compared to the CASPAR slope. The sample of \citet{Hartmann2016} consisted of 148 objects whose accretion rates were computed from continuum excess and line emission with ages of $5<\log t/\mathrm{year}<6$.  This smaller sample size and age range could account for the difference in fit.     

As a simple test of the effect of age, we begin by assuming that objects, no matter their age, share the same $\dot M \sim M^{2.02}$ slope. For older objects, the accretion rates should be lower resulting from a decrease in available disk material.  We can model this simplified assumption as a decrease in the intercept of the $\log \dot M-\log M$ relation with increasing age. This allows us to quantify the extent to which age affects the dispersion in the $\dot M-M$ relation.

\begin{figure}[htp!]
    \centering
    \includegraphics[width=\linewidth]{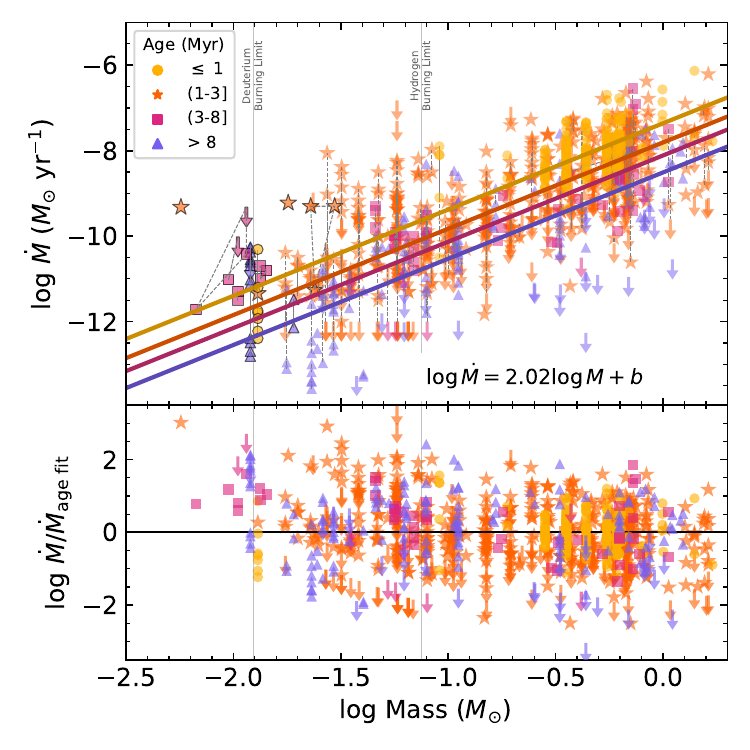}
    \caption{\textit{top:} CASPAR accretion rate vs mass colored by ages: $\leq$1 Myr (gold circles), (1-3] Myr (orange stars), (3-8] Myr (magenta squares), $>$8 Myr (purple triangles). The solid lines are the best linear fits to each age population with a fixed constant slope of 2.02. We find a 1 dex decrease in the best-fit intercept with increasing age. \textit{bottom:} Residuals for the best linear fit for each age population.      }
 \label{fig:MMdotAge}
\end{figure}

As shown in Fig.~\ref{fig:MMdotAge}, we find that the intercept decreases by 1 dex from $<1$ Myr to $>8$ Myr. 
Subtracting age best-fits from each age population and over plotting the residuals gives an indication of the effect of age on overall scatter. If scatter in $\dot M$ for objects of the same mass is a result of age, then we expect the overall dispersion to decrease when age is accounted for in this way. However, we find from these residuals that the $1\sigma$ scatter only decreases by 0.06 dex (bottom panel of Fig.~\ref{fig:MMdotAge}).  
While there is almost 1 dex decrease in best fit intercepts by age bin, the median scatter remains high.  Therefore, as the dispersion within each age bin is roughly the same as the overall scatter, this normalization by age has an insignificant effect.

In the substellar regime, age has an even less effect on the overall scatter; we find the standard deviation of the residuals decreases by 7\%. From Fig.~\ref{fig:new_MMdotResiduals}, the residuals in the substellar regime are positively skewed, exacerbating this effect in the age $\dot M-M$ residuals, especially for the $<1$ and (3-8] Myr regimes.  This could indicate that substellar objects are following a different trend with mass or age.  We investigate this hypothesis in Sec~\ref{sec:BDpop}.

\subsection{Multi-epoch variability}
Previous surveys \citep{Biazzo2014, Costigan2014, Venuti2014} have primarily focused on day-to-day intrinsic accretion variability utilizing one accretion diagnostic. This variability produces $\sim$0.4 dex dispersion, smaller than the 1--2 dex observed in the $\dot M-M$ relation.  The variability is therefore not the dominant source of scatter in the $\dot M-M$ relation for a given mass.   
However, CASPAR contains objects that have been observed from many different tracers across months and years. In this section, we quantify how this longer timescale and methodological accretion variability affects the dispersion we see in the CASPAR $\dot M-M$ relation.

\begin{figure}[htp!]
    \centering
    \includegraphics[width=0.88\linewidth]{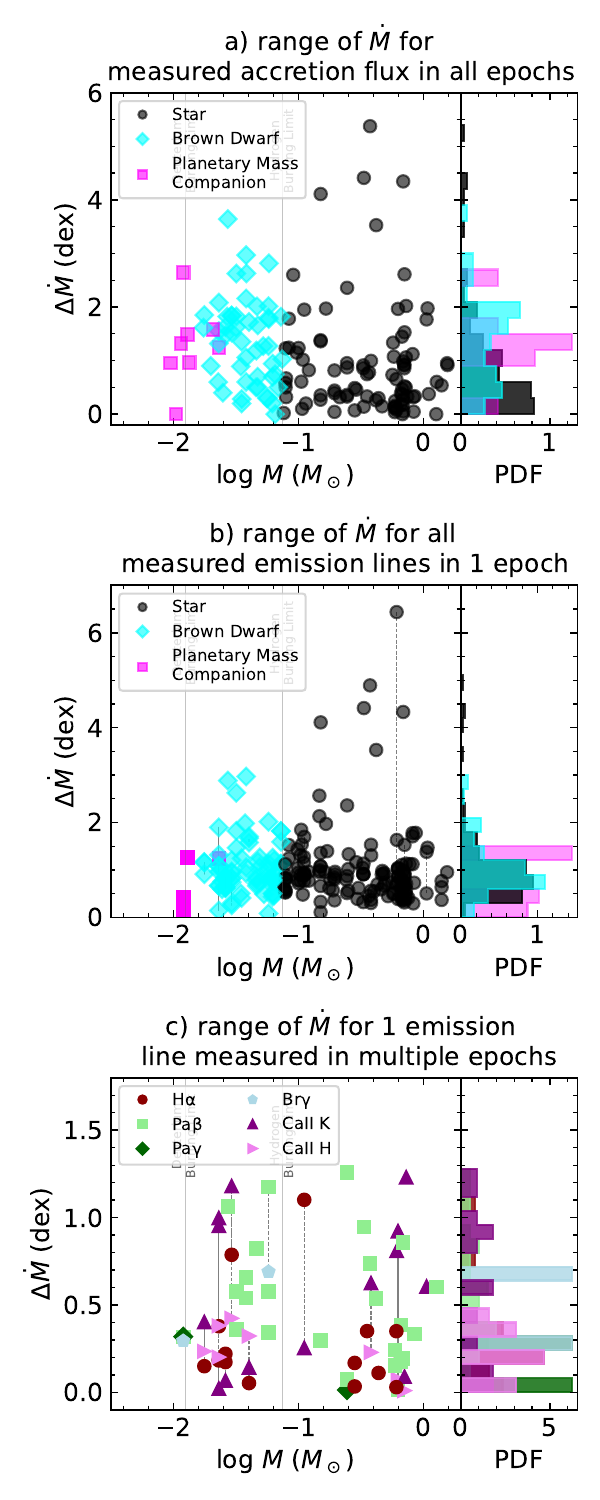}
    \caption{Range of measured accretion rates for objects in CASPAR with a) multi-epoch/multi accretion tracer measurements, b) all accretion rates measured for all emission lines in one epoch, and c) all accretion rates in all epochs for individual emission lines.  In panels a) and b), the ranges are colored by stellar/substellar populations.  In panel c), the ranges are colored by the individual line tracers. 
    The right panels shows the probability density functions for each population. Overall, we find increased variability in the substellar regime (1.33 dex), but consistent $0.8$ dex variability with population for emission lines and epochs.}
    \label{fig:variability}
    \vspace{-10pt} 
\end{figure}

We first look at \textit{all} sources of variability, including variation in accretion rates from different lines in the same epoch, and variation in accretion rate from the same line over multiple epochs. The median separation in measurements is 3 years, with a maximum separation of 21 years from \citet{Gullbring1998} to \citet{Alcala2021}. Overall, we find a spread of 0.63 dex in the stellar regime that includes several high variability outliers from multi-line measurements from a single epoch (see Fig.~\ref{fig:variability}a), and that increases in the substellar regime to a median of 1.33 dex.  We compute an independent two-sample student's t-test, and find a significant (t(134)$=2.08$, $p=0.03$) difference in variability between the stellar (mean $=1.00$ dex, $\sigma=1.21$ dex) and substellar (mean $=1.35$ dex, $\sigma=0.78$ dex) populations. This increase in variability could be indicative of either intrinsic variation resulting from rotation of sunspots or accretion flows in the line-of-sight or differences in diagnostics.     

We also analyze two main sources of observed variation separately, namely: a) variation among accretion rates measured from multiple lines at one observational epoch (``methodological"; 241 objects, Fig.~\ref{fig:variability}b), and b) variation among accretion rates for one line observed at multiple epochs (``intrinsic"; 67 objects, Fig.~\ref{fig:variability}c). See Table~\ref{tab:variability} for the full breakdown of objects.
We find both the median methodological and intrinsic variation in accretion rate is on average 0.86 dex, double the amount of variability found in previous surveys.  The methodological variability in $\dot M$ determination is relatively consistent with mass, however, we do note some strongly variable outliers. We find these objects show small ($< 1.5-2$ dex) variability over multi-decade timespans using a single accretion tracer, but have large ($>3$ dex) variation between accretion rates measured for different lines, with the largest outliers found in the Ophiuchus star forming region ($3.5-5.5$ dex). The object with the most extreme multi-line variability is DO Tau, which has a relatively stable accretion rate of $10^{-8}\ M_\odot$ yr$^{-1}$.  However, \citet{Alcala2021} found an H$\alpha$ line flux of $1.92(\pm0.07)$e$-17$ erg s$^{-1}$ cm$^{-2}$, corresponding to an accretion rate of $6.45\times 10^{-15}\ M_\odot$ yr$^{-1}$ in CASPAR, a $>7$ dex difference in $\dot M$.    

For intrinsic variation, we find Pa$\beta$ and Ca~\textsc{ii} K produce the highest amounts of variability (1.26 dex and 1.23 dex, respectively), though small samples sizes affect the measured ranges of variability.  For Pa$\beta$, we find the median variability increases from stellar (0.33 dex) to substellar (0.57 dex) regimes, though this difference is not statistically significant ($t= -1.42,\ p=0.16$), and as shown by \citet{Claes2022}, measurements found from line luminosity can underestimate the variability for CTTS.

\begin{deluxetable}{lccc}
\tablecaption{Multi-epoch line variability\label{tab:variability}}
\tablewidth{0pt}
\tablehead{\colhead{Line} & \colhead{\# obj} & \colhead{\# multi-epoch} & \colhead{$\Delta \dot M$}}
\startdata
H$\alpha$	& 374	& 16 &	1.10 \\ 
Pa$\beta$	& 207	& 25 &  1.26 \\ 
Pa$\gamma$	& 141	& 2	 &  0.32  \\ 
Br$\gamma$	& 99	& 2	 &  0.69 \\ 
Ca~\textsc{ii} K	& 151	& 14 &	1.23 \\ 
Ca~\textsc{ii} H	& 103	& 8	 & 0.42 \\ 
\enddata
\end{deluxetable}

Though no significant mass trends were seen for either methodological or intrinsic variation, considered separately, when all sources of variation are considered together, we find a significant increase in the variability of $\dot M$ in the substellar regime. As we show in Fig.~\ref{fig:binbyad}b and \ref{fig:byline}, while excess continuum accretion measures remain relatively constant with mass, accretion rates derived from line fluxes, particularly Pa$\beta$, appear to deviate more from the $\dot M-M$ relation as mass decreases.

For objects derived from both excess continuum and line flux, there could be as much as 0.9 dex difference with $\dot M$ estimates. This may point to physical differences in accretion processes in the substellar population (e.g. more luminosity in certain emission lines compared to the expectation from stars) that are not accounted for in current $L_\mathrm{line}\sim L_\mathrm{acc}$ scaling relations. This could have a large impact on our interpretation of ongoing accretion and variability in the lowest mass objects (Betti et al., in prep).       

\section{Role of Accretion on substellar formation}\label{sec:BDpop}
In Sections~\ref{acc diag systematics} and \ref{AllAccretionAge}, we find offsets and skewed residuals in $\dot M-M$ relation for the substellar regime according to age, line tracers, as well as from the overall best linear fit, which might point to underlying differences in the $\dot M-M$ relations needed to describe the stellar and substellar populations.
Therefore, we explore whether the accretion rates of brown dwarfs are statistically distinct from stars and whether this can be connected to their accretion or formation mechanisms. If BDs are accreting differently, this could appear as differences in their calculated accretion rates.    

\begin{figure*}[htp!]
    \centering
    \includegraphics[width=\linewidth]{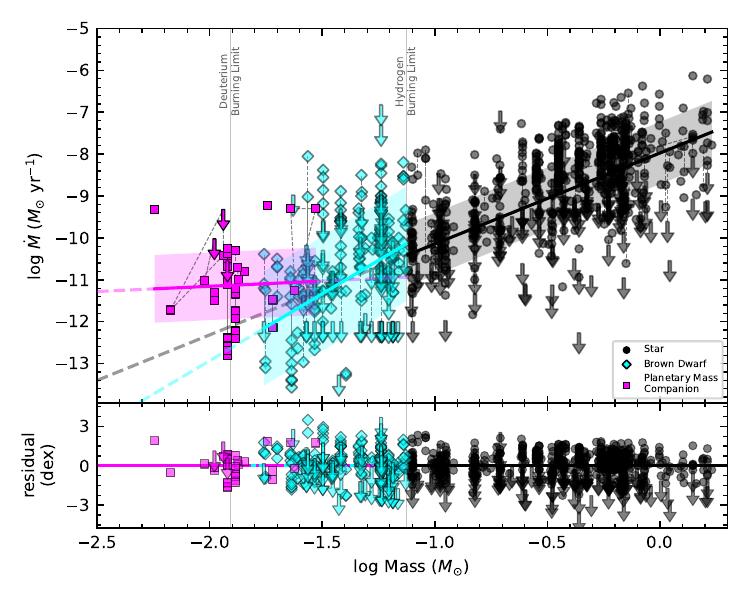}
    \caption{CASPAR $\dot M-M$ best fit relations for stars (black circles), brown dwarfs (cyan diamonds), and planetary mass companions (magenta squares).  The solid lines and bands show the best linear fits and $1\sigma$ standard deviation of the fit for each mass population, while the dashed lines show extrapolations beyond the bounds of the fit regions.  The bottom panel shows the residuals for each best fit. Downward arrows indicate accretion rate upper limits.}
    \label{fig:MMdot_massregions}
\end{figure*}

When we fit each population (stars, brown dwarfs, and planetary mass companions; hereafter mass populations) separately, we find three distinct relationships, shown in Fig.~\ref{fig:MMdot_massregions} and described by the following best-fit relations (hereafter three population fit). For the stellar population, we find:
\begin{equation}
    \log \dot{M} = 2.18(\pm0.11)\log M - 7.97(\pm0.05),
\end{equation}
with a residual standard deviation of 0.74 dex. The accretion rate, $\dot M$, is in $M_\odot$ yr$^{-1}$ and $M$ in $M_\odot$.
For the brown dwarfs:
\begin{equation}
    \log \dot{M} = 3.19(\pm0.57)\log M - 6.54(\pm0.80),
\end{equation}
with a residual standard deviation of 1.36 dex.
Finally, for the planetary mass companions, we find:
\begin{equation}
    \log \dot{M} = 0.25(\pm2.91)\log M - 10.66(\pm5.64),
\end{equation}
with a residual standard deviation of 0.81 dex.
The stellar fit ($\mathrm{slope}=2.18$) is similar to those found by other authors \citep[e.g.][]{Calvet2004, Natta2004, Mohanty2005, Muzerolle2005, Herczeg2008, Zhou2014, Hartmann2016}, as discussed in Section~\ref{scatter1}.

For BDs, the slope of the relation steepens to 3.19 in the substellar regime, while the PMCs are best modeled by a flat dependence with mass.  In Appendix~\ref{app:popstats}, we compare the three population fits to the single population best fit described in Section~\ref{scatter1}, and find that separate fits are statistically favored over the single population fit with lower AIC statistics.

This steeper slope (``knee") in the substellar/low mass star regime has only been seen observationally in older individual star forming systems \citep{Alcala2017, Manara2017b}. As CASPAR includes multiple SFRs at different ages, this steeper BD slope could be a result of these older ages having a stronger effect on the best fit $\dot M-M$ relation for this mass regime. Therefore, in the following sections, we explore the effects of mass with age and mass population in explaining both the scatter and evolution of BD accretion.

\subsection{Effect of Age on Brown Dwarf Properties} \label{age_regions}
\begin{figure}[htp!]
    \centering
    \includegraphics[width=\linewidth]{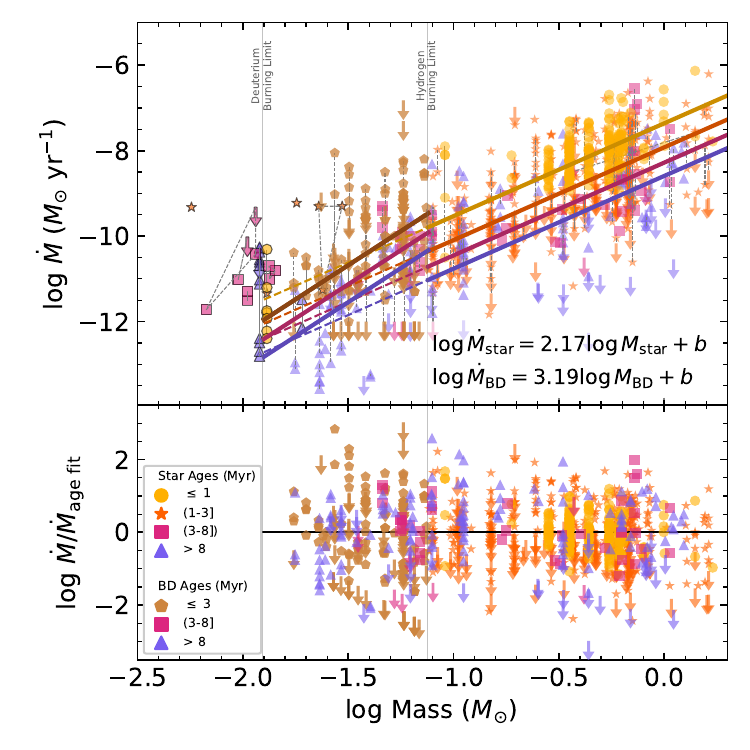}
    \caption{\textit{top:} CASPAR accretion rates vs mass colored by age with best linear fits for each age and mass population (brown dwarf, stars). We combined the brown dwarf $\leq1$ and $(1-3]$ Myr populations to improve statistical number counts (brown pentagons). We extend the stellar fits into the BD regime indicated by dashed lines.  \textit{bottom:} Residuals for the best linear fits for each age and mass populations. The colors and markers are as in Fig.~\ref{fig:MMdotAge}.}
    \label{fig:MMdotAge_regions}
\end{figure}

Following the process outlined in Section~\ref{AllAccretionAge}, we explore the effect of age in explaining the $\dot M-M$ relation for substellar objects.  We combine the original $<1$ Myr and (1-3] Myr populations due to small numbers within the $<1$ Myr bin.  As shown in Fig.~\ref{fig:MMdotAge_regions}, we find consistent decreases in the best-fit y intercept with age for the substellar regime (as well as the stellar regime). Residuals of these age fits show a significant decrease in dispersion as a result of fitting the BD and stellar populations separately; the $1\sigma$ standard deviation of the residuals for the BD population is 1.02 dex. We find a decrease of $\sim$33\% from the 1.36 dex found in Section~\ref{sec:BDpop}. 

In the substellar regime, we find higher accretion rates at the HBL compared to what is expected from the stellar best-fit for each age.  However with the steeper slope, by the deuterium burning limit (DBL), these relationships predict accretion rates just slightly lower than those extrapolated from the stellar best-fits.  Higher mass BDs additionally appear to accrete faster at older ages compared to very low mass stars at similar ages, as seen at the HBL in Fig.~\ref{fig:MMdotAge_regions}.    

\begin{figure}[htp!]
    \centering
    \includegraphics[width=\linewidth]{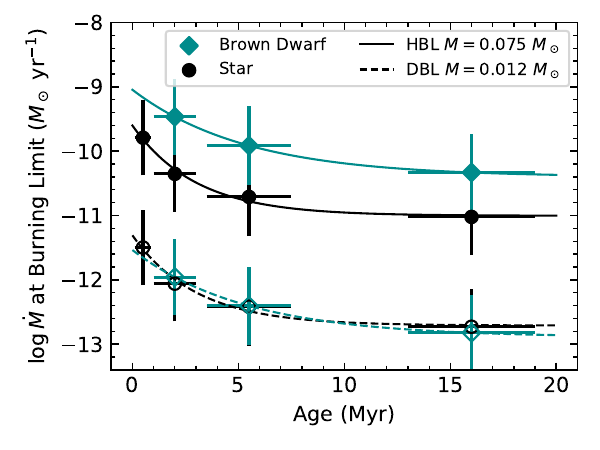}
    \caption{Best fit $\log \dot M$ as a function of age at the hydrogen burning limit ($M=0.075\ M_\odot$; solid line and filled markers) and at the deuterium burning limit ($M=0.012\ M_\odot$; dashed line and open markers) for stars (black circles) and brown dwarfs (cyan diamonds) from Fig.~\ref{fig:MMdotAge_regions}. We show the best exponential fits for each population and find the brown dwarf accretion rates decrease at a faster rate compared to the stellar.}
    \label{fig:interceptage}
\end{figure}

In order to access the rate of decay of accretion for the substellar population compared to the stellar population, we extract $\log \dot M$ at the HBL and DBL for the stellar and substellar fits.
We find the following exponential best fits to these data:
\begin{equation}
    \dot M_\mathrm{stars} \propto 1.40 e^{-t/3.0\ \mathrm{Myr}}
\end{equation}
\begin{equation}
    \dot M_\mathrm{BDs} \propto 1.35 e^{-t/5.2\ \mathrm{Myr}},
\end{equation}
where $\dot M$ is in $M_\odot$ yr$^{-1}$.  We show these fits at the HBL and DBL as a function of binned age in Fig.~\ref{fig:interceptage}. 

The exponential trend in the stellar regime is similar to the exponential disk fraction decay timescale ($\tau=2.5$ Myr) found by \citet{Mamajek2009}, while the decay rate for BD accretion remains high compared to the $\tau=3$ Myr timescale for disk fraction decay \citep{Mamajek2009}, with objects still accreting quickly at older ages.

\subsection{Relationship between Accretion Rate and Mass, Age, \& Mass Population}\label{sec:interaction}
To test whether the relations among mass accretion, object mass, and age show evolutionary trends that could explain the (lack of) knee in the accretion-rate timescale, we fit the $\dot M-M$ relation with age ($t$) and mass as free parameters. More specifically, we fit the model
\begin{equation}
    \log \dot M =  \alpha M + \beta t + \gamma (\log M \times t) + \delta,
\end{equation}
for each mass population, where $\dot{M}$ is in $M_\odot$ yr$^{-1}$, $M$ in $M_\odot$, and $t$ in Myr.

\begin{figure*}[htp!]
    \centering
\includegraphics[width=.8\linewidth]{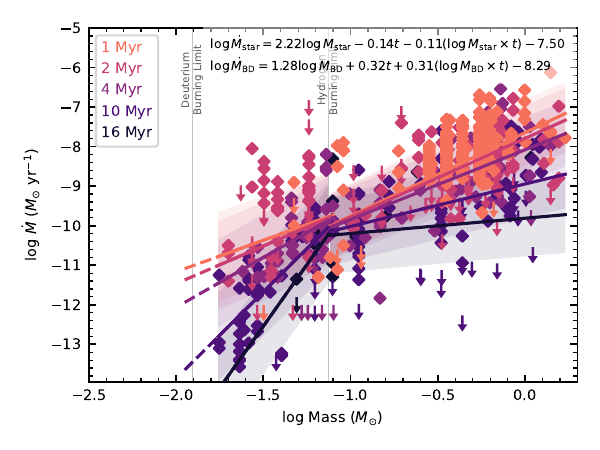}
    \caption{Accretion rate vs mass for the star and brown dwarf population colored by specific ages.  The solid lines and bands show the best linear fits and $1\sigma$ standard deviation where age and mass are both free parameters for each mass population. The dashed lines are extrapolations to the DBL.}
    \label{fig:interactionmodel}
\end{figure*}
We find best fit values for the stellar and BD populations, given by:
\begin{equation}
\begin{split}
    \log \dot M_\mathrm{star} =& 2.22(\pm 0.10)\log M_\mathrm{star} - 0.14(\pm 0.01)t \\
    & - 0.11(\pm0.02)(\log M_\mathrm{star}\times t) - 7.50(\pm0.06)
\end{split}
\end{equation}
and 
\begin{equation}
\begin{split}
    \log \dot M_\mathrm{BD} =& 1.28(\pm 0.60)\log M_\mathrm{BD} - 0.32(\pm 0.14)t \\
    & - 0.31(\pm0.09)(\log M_\mathrm{BD}\times t) - 8.29(\pm0.81)
\end{split}
\end{equation}

As shown in Fig.~\ref{fig:interactionmodel}, in the substellar regime, we find a clear mass and age dependence on accretion rate, with slope increasing with age, while this trend flattens in the stellar regime. In older systems, we see a systematic steepening in the $\dot M-M$ slope compared to the stellar regime. This is suggestive of a faster evolutionary timescale for accretion onto brown dwarfs (especially low mass BDs), as they accrete material at a relatively higher rate ($\dot M \sim 10^{-11}$) at young ages depleting their disks quickly. Higher mass BDs and stars instead accrete for longer at a high accretion rate ($\dot M \sim 10^{-7}-10^{-10}$), indicative of relatively slower disk depletion at younger ages. This confirms previous work in individual SFRs showing a shallower substellar trend for younger systems \citep{Manara2015, Fiorellino2021} and a steeper trend in older systems \citep{Alcala2017, Manara2017b}. 

When we fit to both mass and age, the standard deviation of the best fit residuals for all objects is 0.78 dex, a $9\%$ decrease from the single population best fit standard deviation (0.85 dex).  
Compared to the three population fits, the residual scatter decreases by 6\% (0.71 dex from 0.75 dex) and 44\% (0.94 dex from 1.36 dex) in the stellar and substellar regime.  When we remove the outlier upper limits, these decrease by 17\%, 22\%, and 58\%, for the total, stellar, and brown dwarf populations respectively, indicating both mass and age have a profound effect on the rate of accretion for BDs.  This effect is also apparent when we also bin by age (as discussed above), and have slope as a free parameter (see By Age and Mass in Table~\ref{tab:best fit}).  For the stellar fits, the slope becomes shallower with age, while the BD fits become steeper.  We also see this trend when we examine individual star forming regions in Appendix~\ref{app:SFR}.

\section{Discussion}\label{sec:discuss}
In Section~\ref{scatter1}, we find that uniformly deriving physical and accretion properties reduces the scatter in the CASPAR $\dot M-M$ relation by 17\%, indicating that methodology plays a role in the observed scatter when comparing accretion rates across multiple detection and analysis techniques.  However, since a large ($\sim$1 dex) scatter in the $\dot M-M$ relationship has been seen even in surveys using one detection technique \citep[e.g.][]{Muzerolle2001, Herczeg2008, Alcala2017}, methodology cannot fully explain the observed dispersion around the relation.

We first investigate the roles of accretion diagnostic, intrinsic variability, and age on the $\dot M-M$ scatter.  In the stellar regime, the scatter is consistent among diagnostics, especially between line luminosity and continuum excess.  This is expected as the $L_\mathrm{acc}-L_\mathrm{line}$ relationships are derived for young stellar objects \citep{Alcala2014, Alcala2017, Rigliaco2011}.  We do start to see an increase in the line luminosity scatter in the substellar regime, with the median $\dot M$ increasing $\sim 0.8$ dex off the single population $\dot M-M$ relation (Fig.~\ref{fig:binbyad}), likely driven by NIR emission (Fig.~\ref{fig:byline}). Though we do see intrinsic accretion variability in $\dot M$ over multiple lines at the same epoch or over multiple epochs ($\sim 0.8$ dex), as only 241 and 67 objects in CASPAR are contributing to this variability, respectively, they are likely not driving the scatter.  Indeed, this spread is in line with previous surveys of accretion variability, and smaller than the total observed scatter \citep[see][and references therein]{Manara2022}. Finally, when we just consider age as a driving factor, we find that the $\dot M$ scatter in each age bin is roughly the same as the overall scatter (see Figs.~\ref{fig:MMdotAge} and \ref{fig:MMdotAge_regions}) and normalizing by age has little effect.        

Instead, we posit that the measured accretion rates, dispersion, and variability behavior of BDs is distinct from the stellar regime.  We find that three population (stars, BDs, PMCs) fits are statistically favored over fits to a single population for objects of all masses, and that fitting mass populations separately results in the greatest reduction in residuals. This is most clearly seen in the brown dwarf regime, where fitting for both mass and age reduces the residual scatter by 44-58\%.  
When the substellar and stellar populations are fitted separately, age effects are compensated for, and a uniform methodology is applied to derive accretion rates, the total scatter in the $\dot M-M$ relation across all mass regimes decreases from 1.0 dex to 0.78 dex, a decrease of 28\%.  

The hypothesis of a different $\dot M-M$ relation in the substellar regime is not unique to this work.  \citet{Vorobyov2009} predicted a bi-modality in the $\dot M-M$ relation; in particular, they predicted a steepening at lower masses.  This $\dot M-M$ bi-modality was verified observationally by \citet{Alcala2017} and \citet{Manara2017b} in the Chamaeleon I (1.5 Myr) and Lupus (2.5 Myr) regions, with a break at $M\sim0.2\ M_\odot$. However, this break was not seen in other regions of similar or younger ages \citep[probing the stellar and high mass substellar regime][]{Manara2015, Fiorellino2021}.  \citet{Manara2022} suggests that it could be an evolutionary effect, wherein low mass stars accrete their disk mass at a faster rate during the late stages of formation.  From Fig.~\ref{fig:interactionmodel}, we see a similar trend, where the slope steepens with age for substellar masses, showing that this evolutionary effect is universal for BDs even in different star forming regions.

Theoretical studies of brown dwarfs suggest a variety of formation mechanisms, from more planetary processes such as disk fragmentation \citep[e.g.,][]{Bate2002, Bate2003} to protostellar embryo ejection \citep[e.g.,][]{Goodwin2005, Hubber2005} and turbulent fragmentation \citep[e.g.,][]{Kirk2006}. 
Previous observational surveys of brown dwarfs have found that their disk properties follow trends similar to stars, with similar disk fractions \citep{Luhman2005}, correlations between disk and stellar mass \citep{Testi2016, WardDuong2018, Sanchis2021, Rilinger2021}, and an inverse correlation between disk mass and age \citep{Rilinger2021}.  The lowest mass BDs should have small disks; however, at the youngest ages, they appear to accrete material at similar rates to higher mass BDs with larger disks (within an order of magnitude; see Fig.~\ref{fig:interactionmodel}). In other words, the slope of the $\dot M-M$ relation for the youngest brown dwarfs is shallow.  The steepening of the relation in $\dot M-M$ with age could be a result of the lowest mass brown dwarfs accreting ``too rapidly" at younger ages and depleting their disk material.

Combining both physical and systematic offsets in the BD regime, evidence from CASPAR points to a two-fold complication when deriving accretion rates for the lowest mass objects. 
First, as discussed above, age and mass play a significant role in accretion rates (see Fig.~\ref{fig:interactionmodel}). Additionally, accretion rates derived from line luminosities are calculated using empirically derived $L_\mathrm{acc}-L_\mathrm{line}$ scaling relationships for stars. However, if stellar and substellar objects follow different $\dot M-M$ relationships, there is no guarantee that the same $L_\mathrm{acc}-L_\mathrm{line}$ relationships accurately represent single (or bound) brown dwarfs.  By using stellar relationships in the substellar regime, we could be artificially overestimating both accretion luminosities and accretion rates derived from line luminosities. Systematic variations in accretion rate scalings may play a role in the apparent variation among diagnostics for substellar objects, with up to 0.8 dex difference in inferred accretion rates for certain tracers (e.g., line fluxes vs. excess continuum).  
Forthcoming work will explore this issue (Betti et al., in prep), which may have a profound impact on the interpretation of substellar accretion. 

\subsection{Planetary Mass Companion Population}
In this section, we describe briefly trends seen in the PMC population. Though the sample of accreting PMCs is small and has not been rederived consistently in mass or accretion rate, we note that these objects appear to follow a much flatter and higher relation compared to BDs.  \citet{Stamatellos2015} modeled disk and accretion properties of bound companions formed via disk fragmentation, allowing a gravitationally unstable disk with a mass of 0.7 $M_\odot$ around a 0.7 $M_\odot$ star to grow until it started to fragment. They predict that the disks around PMCs are more massive than expected for objects of the same mass forming in isolation from a collapsing core.  This is due to the fact that PMCs are forming within the larger circumstellar disk. Before they separate from the disk (become dynamically independent), they are able to accrete more gas than a BD whose only material reservoir is its natal core \citep{Stamatellos2015}. \citet{Stamatellos2015} predict no strong correlation between object mass and disk mass under this scenario, leading to a flatter slope in $\dot M$ vs. $M$.   

The difference that we observe between high PMC accretion rates and low accretion rates for brown dwarfs of similar mass could be either: a) an observational bias in observed PMCs, such that objects with accretion rates below $10^{-12}\ M_\odot$ yr$^{-1}$ have simply gone undetected, b) PMCs may be fundamentally different from BDs, for example by forming via a disk fragmentation-like mechanism as opposed to core collapse, or c) using incorrect scaling relations pushes our $\dot M$ estimates of all substellar objects up or increases the dispersion in the relation.

Recent theoretical work by \citet{Aoyama2018}, \citet{Aoyama2020}, and \citet{Marleau2022a} also show that bound PMCs may have a higher fraction of line emission contributing to their total accretion luminosity than accreting stars. These models predict significantly larger ($\sim$2 dex) accretion rates for planetary-mass objects than those derived from stellar magnetospheric empirical relations \citep{Alcala2014, Alcala2017}, which would drive these $\dot M$s to even higher rates. As BDs are traditionally used to place the accretion of PMCs in context, more care will have to be taken in calculating and interpreting accretion signatures from PMCs.

\section{Conclusion}\label{sec:conclusion}
In this paper, we introduce CASPAR, the Comprehensive Archive of Substellar and Planetary Accretion Rates, the largest database of substellar and planetary accretion rates to-date. The physical and accretion properties in CASPAR have been rederived under a consistent methodology (Gaia distances, consistent evolutionary models and scaling relations).  The goal of this effort was to investigate the contribution of systematic offsets among methods to overall scatter in the $\dot M-M$ relation. Using the rederived database, we investigate the dispersion in the $\dot M-M$ relation, and the physical and systematic processes that contribute to it. We also explore variation among stars, brown dwarfs and planetary mass companion populations.  We find:
\begin{itemize}
    \item  Rederiving all physical and accretion properties using the same methodology decreases the 1.04 dex of scatter about the single population $\dot M-M$ fit by 17\%. The best single population linear fit for CASPAR, $\log \dot M = 2.02\log M - 8.02$, is consistent with previous estimates from smaller samples, 
    suggesting that methodological differences in derivation play a small role in the slope or scatter of the $\dot M-M$ relation.
    \item The choice of accretion diagnostic additionally contributes to the overall scatter at substellar masses, with estimates from line luminosities leading to an average of 0.8 dex variation for a single object. Within the stellar regime, accretion rates are consistent among  tracers.  Unlike line luminosity, excess continuum derived estimates are consistent to within 0.21 dex of the overall best linear fit across both substellar and stellar mass regimes ($0.1 \lesssim M/M_\odot \lesssim 2$).
    \item We also find consistent multi-epoch and multi-tracer variability of $\sim0.6$~dex in the stellar regime, consistent with previous estimations.  This variability increases to 1.33~dex in the substellar regime.  We posit that this increase is due to either higher variability seen at lower masses or stellar scaling relations being invalid in substellar regimes, leading to offsets in derived accretion rates. 
    \item We investigate the effect of age on dispersion around the relation and find a 1 dex decrease in the best fit intercept between $\sim1-10$~Myr.  However, the scatter within age bins is $\sim1$ dex, leading to little change in the overall scatter compared to the scatter from the overall best fit residuals.    
    \item We argue that the majority of the scatter can be explained by modeling the star, brown dwarf, and PMC populations by separate $\dot M-M$ relations, accounting for both mass and age. We find the brown dwarf $\dot M-M$ scatter decreases by 44\% as a result (58\% when upper limit outliers are excluded from the residuals). Additionally, we show that the BD $\dot M-M$ relation steepens with age, while the stellar relation flattens.  
    
    We posit that there is a two-fold issue when deriving accretion rates for low mass objects. First, accretion rates are expected to depend much more on age and mass than in the stellar regime. Secondly, accretion rates derived from stellar scaling relations likely overestimate BD accretion rates, contributing to the overall scatter in this population. 
    \item Bound planetary companions seem to follow a flatter $\dot M-M$ relation compared to brown dwarfs and stars. This may be a result of differences in either their formation or accretion paradigms.  Accretion measurements for a larger population of PMCs and individual modeling of these systems will help reveal the underlying physics governing them.  
\end{itemize}

CASPAR is an evolving database and with future/ongoing surveys \citep[e.g.,][]{Pittman2022, Gangi2022}, protoplanet detections \citep[e.g.,][]{Ringqvist2023}, and derived scaling relations \citep[e.g.,][]{Marleau2022b}, it will continue to be updated. All updates and additions will be found on Zenodo at the following: \url{https://doi.org/10.5281/zenodo.8393054}. Suggestions for additions to CASPAR can be made to the lead author.  
  
\begin{acknowledgments}
We thank the anonymous referee for their careful and thoughtful review that truly benefited this manuscript.
S.\ K.\ Betti acknowledges support for this work from the NASA Future Investigators in NASA Earth and Space Science and Technology grant 80NSSC22K1750 (FI: S. K. Betti; PI: D. Calzetti, co-I: K. Follette).
K. B. Follette acknowledge funding from NSF-AST-2009816 and the Cottrell Scholar Award \#28290.
G.-D.\ Marleau acknowledges the support of the DFG priority program SPP 1992 ``Exploring the Diversity of Extrasolar Planets'' (MA~9185/1),
and from the Swiss National Science Foundation under grant 200021\_204847 ``PlanetsInTime''.
This work has made use of data from the European Space Agency (ESA) mission
{\it Gaia} (\url{https://www.cosmos.esa.int/gaia}), processed by the {\it Gaia}
Data Processing and Analysis Consortium (DPAC,
\url{https://www.cosmos.esa.int/web/gaia/dpac/consortium}). Funding for the DPAC
has been provided by national institutions, in particular the institutions
participating in the {\it Gaia} Multilateral Agreement. This research has made use of the SIMBAD database, operated at CDS, Strasbourg, France. This research has made use of the VizieR catalogue access tool, CDS, Strasbourg, France (DOI: 10.26093/cds/vizier). The original description 
 of the VizieR service was published in 2000, A\&AS 143, 23.  
Parts of this work have been carried out within the framework of the NCCR PlanetS supported by the Swiss National Science Foundation.

\end{acknowledgments}

\vspace{5mm}

\software{astropy \citep{2013A&A...558A..33A,2018AJ....156..123A}, Matplotlib \citep{Hunter2007}, numpy \citep{harris2020array}
 astroquery \citep{astroquery}, scipy \citep{2020SciPy-NMeth}} 
 
\appendix

\section{CASPAR column names}\label{app:CASPAR}
In Table~\ref{tab:CASPAR}, we give all columns within the database and their description.  These are identical between the Literature Database and CASPAR. 
\startlongtable
\begin{deluxetable*}{ll}
\tablecaption{Contents of CASPAR and Literature Database\label{tab:CASPAR}}
\tablewidth{0pt}
\tablehead{\colhead{Column Label} & \colhead{Description, Comment, and/or Units}}
\startdata
Unique ID	&	CASPAR source ID	\\
Unique Name	&	2MASS Point Source Catalog or common name ID	\\
Simbad-Resolvable Name	&	ID resolvable with Simbad	\\
Reference Name	&	Source ID used by reference	\\
Duplicate \#	&	\# for duplicate object	\\
Total Duplicates	&	\# of duplicate objects	\\
Binary	&	Binary flag	\\
Companion	&	Flag on companion	\\
Separation	&	Separation of binary or companion (arcsec)	\\
Object	&	Type of object (star/brown dwarf/PMC)	\\
RA (J2000.0)	&	Right ascension J2000 (deg)	\\
Dec (J2000.0)	&	Declination J2000 (deg)	\\
RA (J2016.0)	&	Gaia Right ascension J2016 (deg)	\\
Dec (J2016.0)	&	Gaia Declination J2016 (deg)	\\
Disk Type	&	Disk class	\\
Association	&	Star forming region or association	\\
Association Probability Banyan Sigma	&	Banyan $\Sigma$ association probability \citep{Gagne2018}	\\
Association Census Reference	&	Association census reference object is in	\\
Association Age	&	Age of association from isochrone fitting (Myr)	\\
Assocation Age err	&	Uncertainty on association age (Myr)	\\
Individual Age	&	Individual object age (Myr)	\\
Individual Age err	&	Individual object age uncertaity (Myr)	\\
Individual Age Reference	&	Reference for individual age	\\
GAIA DR2 Source ID	&		\\
GAIA DR2 Parallax	& (mas)		\\
GAIA DR2 Parallax err	& (mas)		\\
GAIA DR2 Reliable Parallax	& Parallax reliability flag		\\
GAIA DR2 Distance	&	From \citet{Bailer-Jones2018} (pc)	\\
GAIA DR2 Distance lower limit	&	From \citet{Bailer-Jones2018} (pc)	\\
GAIA DR2 Distance upper limit	&	From \citet{Bailer-Jones2018} (pc)	\\
GAIA DR2 RA proper motion	& (mas/yr)		\\
GAIA DR2 RA proper motion err	& (mas/yr)		\\
GAIA DR2 Dec proper motion	& (mas/yr)		\\
GAIA DR2 Dec proper motion err	& (mas/yr)		\\
GAIA EDR3 Source ID	&		\\
GAIA EDR3 Parallax	& (mas)		\\
GAIA EDR3 Parallax err	& (mas)		\\
GAIA EDR3 Reliable Parallax	& Parallax reliability flag		\\
GAIA EDR3 Geometric Distance	&	From \citet{Bailer-Jones2021} (pc)	\\
GAIA EDR3 Geometric Distance lower limit	&	From \citet{Bailer-Jones2021} (pc)	\\
GAIA EDR3 Geometric Distance upper limit	&	From \citet{Bailer-Jones2021} (pc)	\\
GAIA EDR3 RA proper motion	&	(mas/yr)	\\
GAIA EDR3 RA proper motion err	&	(mas/yr)	\\
GAIA EDR3 Dec proper motion	&	(mas/yr)	\\
GAIA EDR3 Dec proper motion err	&	(mas/yr)	\\
radial velocity	(barycentric) &	(km/s)	\\
radial velocity err	(barycentric) &	(km/s)	\\
A\_V	&	Visual extinction (mag)	\\
A\_V err	&	Visual extinction uncertainty (mag)	\\
A\_J	&	$J$-band extinction (mag)	\\
A\_J err	&	$J$-band extinction uncertainty (mag)	\\
A\_V reference	&	Visual extinction reference	\\
Jmag	&	$J$ apparent magnitude (mag)	\\
Jmag err	&	$J$ apparent magnitude uncertainty(mag)	\\
Hmag	&	$H$ apparent magnitude (mag)	\\
Hmag err	&	$H$ apparent magnitude uncertainty(mag)	\\
Kmag	&	$K$ apparent magnitude (mag)	\\
Kmag err	&	$K$ apparent magnitude uncertainty(mag)	\\
Ha mag	&	H$\alpha$ apparent magnitude (mag)	\\
Ha mag err	&	H$\alpha$ apparent magnitude uncertainty(mag)	\\
Reference	&	Original literature source for accretion rate measurement	\\
Telescope/Instrument	& Facility used to measure accretion rate		\\
Association	&	Association/star forming region used by CASPAR	\\
Age	&	CASPAR age (Myr)	\\
Age err	&	Average CASPAR age uncertainty (Myr)	\\
Epoch	&	Epoch of original accretion rate tracer observation	\\
log g	&	Surface gravity	\\
Distance	&	CASPAR distance (pc)	\\
Distance err lower limit	&	CASPAR distance lower sigma (pc)	\\
Distance err upper limit	&	CASPAR distance upper sigma (pc)	\\
Sp Type	&	Spectral type	\\
Sp Type err	&	Spectral type uncertainty	\\
Teff	&	Effective temperature (K)	\\
Teff err	&	Effective temperature uncertainty (K) 	\\
Mass	&	($M_\odot$)	\\
Mass err	&	($M_\odot$)	\\
Luminosity	&	($L_\odot$)	\\
Luminosity Err	&	($L_\odot$)	\\
Radius	&	($R_\odot$)	\\
Radius err	&	($R_\odot$)	\\
Accretion Diagnostic	&	Method to derive $\dot{M}$ 	\\
Tracer	&	Lines or continuum 	\\
H$\alpha$ 10\% Upper Limit	&	Upper limit flag on $\dot{M}$	\\
H$\alpha$ 10\%	&	(km/s)	\\
H$\alpha$ 10\% err	&	(km/s)	\\
H$\alpha$ 10\% Accretion Rate	&	$\dot{M}$ derived from H$\alpha$ 10\%	\\
Line$^{*}$ Upper Limit	&	Upper limit flag on $\dot{M}$ for line	\\
Line$^{*}$ EW	&	Equivalent width for line (\AA)	\\
Line$^{*}$ EW err	&	Equivalent width uncertainty for line (\AA)	\\
Line$^{*}$ Line Flux	&	Line Flux for line (erg/s/cm$^2$) 	\\
Line$^{*}$ Line Flux err	&	Line Flux uncertainty for line (erg/s/cm$^2$) 	\\
Line$^{*}$ Log Accretion Luminosity	&	Accretion Luminosity scaled from line ($L_\odot$)	\\
Line$^{*}$ Log Accretion Luminosity err	&	Accretion Luminosity uncertainty scaled from line ($L_\odot$)	\\
Line$^{*}$ Accretion Rate	&	Accretion Rate for line ($M_\odot /$yr)	\\
Line$^{*}$ Accretion Rate err	&	Accretion Rate uncertainty for line ($M_\odot /$yr)	\\
Upper Limit	&	Flag on $\dot{M}$ upper limit	\\
Log Accretion Luminosity	&	Accretion Luminosity from Accretion Diagnostic ($L_\odot$)	\\
Log Accretion Luminosity err	&	Accretion Luminosity uncertainty from Accretion Diagnostic ($L_\odot$)	\\
Accretion Rate	&	Accretion Rate from Accretion Diagnostic ($M_\odot /$yr)	\\
Accretion Rate err	&	Accretion Rate uncertainty from Accretion Diagnostic ($M_\odot /$yr)	\\
Scaling Relation	&	$L_\mathrm{line}-L_\mathrm{acc}$ scaling relation reference	\\
SpTemp Conversion	&	Spectral type to temperature conversion reference	\\
Evolutionary Models	&	Evolutionary Model method	\\
Notes	&		\\
Links	&	Link to SIMBAD	\\
\enddata
\tablenotetext{*}{Lines: H$\alpha$, H$\beta$, H$\gamma$, H$\delta$, H$\epsilon$, H8, H9, H10, H11, H12, H13, H14, H15, Pa$\beta$, Pa$\gamma$, Pa$\delta$, Pa8, Pa9, Pa10,  Br$\gamma$, Br8 Pf$\beta$, He~\textsc{i} $\lambda$\,4026, He~\textsc{i} $\lambda$\,4471, He~\textsc{i} $\lambda$\,4713, He~\textsc{i} $\lambda$\,5016, He~\textsc{i} $\lambda$\,5876, He~\textsc{i} $\lambda$\,6678, He~\textsc{i} $\lambda$\,7065, He~\textsc{i} $\lambda$\,10830, He~\textsc{ii} $\lambda$\,4686, Ca~\textsc{ii} K, Ca~\textsc{ii} H, Ca~\textsc{ii} $\lambda$\,8498, Ca~\textsc{ii} $\lambda$\,8542, Ca~\textsc{ii} $\lambda$\,8662, Na~\textsc{i} $\lambda$\,5889, Na~\textsc{i} $\lambda$\,5896, O~\textsc{i} $\lambda$\,8446, C~\textsc{iv} $\lambda$\,1549 }
\tablecomments{Only a portion of this table is shown here to demonstrate its form and content.  Machine readable versions of CASPAR (Part~1) and the Literature Database (Part~2) are available.}
\end{deluxetable*}

\section{Literature Database compilation}\label{extra}
\subsection{Kinematic, Photometry, and Age information}\label{sec:kin_phot_age}
For each object, we compiled kinematic information (Right Ascension, Declination, parallax, distance, proper motion, and radial velocity) from Gaia DR2 \citep{GAIADR22016, GAIADR22018} and Gaia EDR3 \citep{GAIAEDR32016, GAIAEDR32021}, in case an object was not observed in one of the data releases.  We queried the Gaia archives in order to find the Gaia object associated with each database entry. Of the 793 unique objects, we retrieved kinematic information for 670 that have Gaia observations (of the substellar object, 65/87 have Gaia observations). If the parallax is considered reliable \citep[e.g., its parallax error is less than 25\% and parallax $>$ 0.167 mas;][]{Huang2022}, we use the geometric distances found for DR2 and EDR3 by \citet{Bailer-Jones2018} and \citet{Bailer-Jones2021}, respectively. Near infrared (NIR) photometry for each object is compiled from 2MASS, while the accretion literature reference H$\alpha$ photometry is included for objects whose accretion rates are measured from this photometry.  

For objects with D $<150$ pc, we ran their kinematic information through the Banyan $\Sigma$ tool \citep{Gagne2018} in order to determine the most likely host association. If the membership probability is $>70$\%, we assume membership in that association and assign the object the corresponding age from Table~\ref{tab:ages}.  For those objects D $>150$ pc or whose membership probabilities were lower than 70\%, we searched the literature for population census studies that may have determined the membership in an association or a cluster.  Finally, for those objects not in Banyan $\Sigma$ or census papers, we determine if a) it is a known field star, or b) if its kinematic information is close to an association.  If the latter, we assign it to the closest association with the caveat that the association is only assumed, indicated by a ``*" next to the assigned association in the database.   

In order to estimate object ages, we compile a list of the star forming regions (SFR), clusters, associations, and clouds associated with each object.Star formation is not instantaneous within molecular clouds, but occurs on individual, sub-group, and regional levels, leading to gradients and even separate age populations within these associations.  Additionally, effects such as extinction, accretion history, and binarity affect the observational uncertainty when determining ages \citep{Krolikowski2021}.  These can lead to vertical scatter on Hertzsprung--Russell diagrams \citep[HRD;][]{Baraffe2017}, a result of the variation in the radius-to-mass ratio driven primarily by variations such as age \citep{Pecaut2012, Soderblom2014, Rizzuto2016}.  These large uncertainties make individual age estimates hard to derive from HRDs and rife with large uncertainties \citep{Pecaut2012, Malo2014, Feiden2016, Rizzuto2020}. Additionally, other methods of calculating ages (i.e. lithium burning, kinematics) all have intrinsic assumptions and systematic effects on the error \citep[see][for a detailed review of these effects]{Soderblom2014}.  

Therefore, recent studies of SFRs have relied on robustly estimating ages of groups within SFRs, which \textit{reduces} the vertical spread resulting from assuming a single age as well as the large uncertainty from assuming individual ages \citep[e.g.][]{Galli2020, Galli2021, Krolikowski2021, Esplin2020}.
This method relies on placing objects on a HRD and comparing to known isochrone and evolutionary track models to estimate age and mass, respectively. Though there are systematic and observational uncertainties due to potential differences in models used, and assumptions of temperature and luminosity, this method has been employed previously for all of the associations found in CASPAR, providing a (semi-)uniform method to derive ages.  We investigate if assuming a single age for the whole SFR affects the resulting masses and accretion rates (and the effect on the $\dot M-M$ scatter). Using Taurus, as there is a large number of subregions and ages, we compare the radius-to-mass ratio derived from assuming a single age ($2\pm1$ Myr) and ages from the separate regions (see Table~\ref{tab:ages}). We find that assuming a single age for a whole SFR only affects the radius/mass ratio and accretion rate by less than $<20\%$ compared to using region ages.  This is generally less than the uncertainty on both individual and SF ages ($\sim$ 50\%) and likely does not greatly affect the derived accretion rates. We therefore use the average age from the most recent census papers for each region (see Table~\ref{tab:ages}) corresponding to the (sub)cluster/association that the object is a member. Individual object ages estimated from isochrone fitting from the literature reference are also included in the database.

\begin{deluxetable*}{lccccc}
\tablecaption{Star forming regions and Association ages and distances\label{tab:ages}}
\tablewidth{0pt}
\tablehead{\colhead{Region} & \colhead{\# in CASPAR} & \colhead{Age}  &  \colhead{Age} & \colhead{Distance} & \colhead{Distance} \vspace{-3mm}\\
\colhead{} & \colhead{} &\colhead{(Myr)}  &  \colhead{Ref.} & \colhead{(pc)} & \colhead{Ref.}}
\startdata
25 Orionis	&	1	&	6.2	$\pm$	2.3	&	1	&	354	$\pm$	3	&	1	\\
118 Tau	&	1	&	10			&	2	&				&		\\
Argus	&	1	&	45	$\pm$	5	&	3	&				&		\\
$\beta$ Pictoris	&	1	&	24	$\pm$	3	&	4	&	40	$\pm$	17	&	5	\\
Chamaeleon I North	&	57	&	1.5	$\pm$	0.5	&	6	&	191	$\pm$ 	0.8	& 	6	\\
Chamaeleon I South	&	47	&   1.5	$\pm$	0.5				&	6	&	187 	$\pm$ 	1	&  6		\\
Corona-Australis	&	3	&	1.5	$\pm$	0.5	&	7	&				&		\\
$\eta$ Chamaeleontis	&	15	&	11	$\pm$	3	&	4	&	94			&	8	\\
IC 348	&	8	&	4	$\pm$	2	&	9	&	321	$\pm$	10	&	10	\\
Lagoon Nebula	&	224	&	0.7	$\pm$	0.4	&	11	&	1326$_{-69}^{+77}$			& 	12	\\
Lupus	&	19	&	2.6	$\pm$	0.5	&	13, 14	&	158	$\pm$	0.6	& 	13	\\
Lupus 1	&	3	&	1.2	$\pm$	0.6	&	13	&	155$_{-3.4}^{+3.2}$			&	13	\\
Lupus 2	&	5	&	2.6	$\pm$	0.5	&	13	&	158$_{-5}^{+7}$			&	13	\\
Lupus 3	&	34	&	2.5	$\pm$	0.5	&	13	&	159	$\pm$	0.7	&	13	\\
Lupus 4	&	7	&	2.4	$\pm$	1.3	&	13	&	160	$\pm$	1	&	13	\\
NGC 2024	&	1	&	1.1	$\pm$	1	&	1	&				&		\\
$\rho$ Oph	&	26	&	$\sim$6			&	15	&	140			&	15	\\
$\rho$ Oph/L1688	&	76	&	2	$\pm$	1.2	&	15	&	138			&	15	\\
$\sigma$ Ori	&	89	&	1.9	$\pm$	1.6	&	1	&	406	$\pm$	4	&	1	\\
Sh 2-284	&	3	&	3.5	$\pm$	1	&	16	&	4000			&	16	\\
Taurus	&	12	&	2	$\pm$	1	&	17	&	140			&	17	\\
Taurus/B213	&	2	&	3.1	$\pm$	0.9	&	17	&	156			&	17	\\
Taurus/L1495	&	13	&	1.3	$\pm$	0.2	&	17	&	130			&	17	\\
Taurus/L1517-Center	&	6	&	2.5	$\pm$	1	&	17	&	155			&	17	\\
Taurus/L1517-Halo	&	1	&	2.3	$\pm$	0.5	&	17	&	157			&	17	\\
Taurus/L1524	&	15	&	1.6	$\pm$	0.9	&	17	&	128			&	17	\\
Taurus/L1527	&	5	&	2.6	$\pm$	0.8	&	17	&	142			&	17	\\
Taurus/L1544	&	1	&	3.4	$\pm$	0.9	&	17	&	168			&	17	\\
Taurus/L1546	&	5	&	2	$\pm$	0.3	&	17	&	160			&	17	\\
Taurus/L1551	&	8	&	1.7	$\pm$	0.2	&	17	&	145			&	17	\\
Taurus/L1558	&	3	&	3.3	$\pm$	0.4	&	17	&	147			&	17	\\
Taurus/North	&	5	&	2.5	$\pm$	0.4	&	17	&	143			&	17	\\
Taurus/South	&	1	&	6.2	$\pm$	1.7	&	17	&	123			&	17	\\
Tucana-Horologium	&	1	&	45	$\pm$	4	&	4	&				&		\\
TW Hydra	&	13	&	10	$\pm$	3	&	4	&	50			&	18	\\
Upper Centaurus Lupus	&	15	&	16	$\pm$	2	&	19	&	130$_{-32}^{+62}$			&	19	\\
Upper Scorpius	&	68	&	10	$\pm$	3	&	14, 19, 20	&	141$_{-37}^{+77}$			& 	19	\\
\enddata
\tablerefs{(1) \citet{Kounkel2018}, (2) \citet{Gagne2018}, (3) \citet{Zuckerman2019}, (4) \citet{Bell2015}, (5) \citet{Messina2017}, (6) \citet{Galli2021}, (7) \citet{Esplin2022}, (8) \citet{vanLeeuwen2007}, (9) \citet{Bell2013}, (10) \citet{Ortiz-Leon2018}, (11) \citet{Prisinzano2019}, (12) \citet{Wright2019}, (13) \citet{Galli2020}, (14) \citet{Luhman2020}, (15) \citet{Esplin2020}, (16)  \citet{Kalari2015b}, (17)  \citet{Krolikowski2021}, (18)  \citet{Schneider2016}, (19) \citet{Pecaut2016},
(20) \citet{Ratzenbock2023}}
\end{deluxetable*}

\subsection{Literature Database}\label{lit compilation}
For each accretion rate, we compile the literature reference, and physical and accretion properties used to calculate $\dot M$.  This includes the association, age, and distance specific to the literature reference, as well as the spectral type, mass, luminosity and radius.  If any quantity is not specified in its reference, it is left blank.  We also record, if given, the reference used to convert from spectral type to temperature, the evolutionary model used to derive estimates of physical parameters, and the emission line luminosity to accretion luminosity ($L_\mathrm{line}-L_\mathrm{acc}$) scaling relation. 
For each accretion rate, we list the specific accretion signature or continuum band used to calculate $\dot M$.  We list the main accretion diagnostic for each reference in Table~\ref{tab:litrefs2}. 
We also compile all individual line emission quantities (EWs, line fluxes, accretion luminosities, and accretion rates) reported in the paper. In cases where the accretion rate is not presented in the reference, but can be calculated, we have done so and included them in the database.  Finally, the accretion luminosity and rate found from the main reported literature reference are recorded. Objects with multiple measurements (from multiple studies, tracers, or epochs) are connected with a gray dashed line, and highlight the variations in published estimates of mass and mass accretion rate across multiple studies and accretion diagnostics. 

\begin{deluxetable*}{lll}								
\tablecaption{Literature References\label{tab:litrefs2}}					
\tablewidth{0pt}					
\tablehead{\colhead{Reference}  & \colhead{Accretion Diagnostic} & \colhead{Tracer(s) or spectra wavelength range}}					
\startdata					
\citet{Alcala2014}	&	Continuum Excess	&	$\approx3400$--3600, $\approx4000$--4750, 3600, 4600, 7100 \AA	\\
\citet{Alcala2017}	&	Continuum Excess	&	$\approx3400$--3600, $\approx4000$--4750, 3600, 4600, 7100 \AA	\\
\citet{Alcala2019}	&	Line Luminosity	&	C~\textsc{iv} $\lambda$\,1549	\\
\citet{Alcala2020}	&	Continuum Excess	&	$\approx3400$--3600, $\approx4000$--4750, 3600, 4600, 7100 \AA	\\
\citet{Alcala2021}	&	Line Luminosity	&	H$\alpha$, H$\beta$, H$\gamma$, H$\delta$, Pa$\beta$, Pa$\gamma$, Pa$\delta$, Pa$\epsilon$,	\\
	&		&	Ca~\textsc{i} $\lambda$\,3934, He~\textsc{i} $\lambda$\,4026, $\lambda$\,4471, $\lambda$\,4713, $\lambda$\,4922,	\\
	&		&	He~\textsc{i} $\lambda$\,5016, $\lambda$\,5876, $\lambda$\,6679, $\lambda$\,10830	\\
\citet{Betti2022b}	&	Line Luminosity	&	Pa$\gamma$, Pa$\beta$, Br$\gamma$	\\
\citet{Bowler2011}	&	Line Luminosity	&	Pa$\beta$	\\
\citet{Close2014}	&	Line Luminosity	&	H$\alpha$ photometry	\\
\citet{Comeron2010}	&	Line Luminosity	&	Ca~\textsc{ii}	\\
\citet{Eriksson2020}	&	Line Luminosity	&	H$\alpha$ 10\% width, H$\alpha$, H$\beta$, He~\textsc{i} $\lambda$\,6678	\\
\citet{Espaillat2008}	&	Line Profile	&	H$\alpha$	\\
\citet{Gatti2006}	&	Line Luminosity	&	Pa$\beta$	\\
\citet{Gatti2008}	&	Line Luminosity	&	Pa$\gamma$	\\
\citet{Gullbring1998}, 	&	Continuum Excess	&	3200--5400 \AA	\\
   \;\;\citet{Calvet1998}	&	&		\\
\citet{Haffert2019}	&	Line Profile	&	H$\alpha$ 10\% width	\\
\citet{Hashimoto2020}	&	Line Luminosity	&	H$\alpha$, H$\beta$	\\
\citet{Herczeg2008}	&	Continuum Excess	&	3200--9000 \AA	\\
\citet{Herczeg2009}	&	Continuum Excess	&	3200--9000 \AA	\\
\citet{Ingleby2013}	&	Continuum Excess	&	1570--7000 \AA	\\
\citet{Kalari2015}	&	Line Luminosity	&	H$\alpha$ photometry	\\
\citet{Kalari2015b}	&	Line Luminosity	&	H$\alpha$ photometry	\\
\citet{Lee2020}	&	Line Profile	&	H$\alpha$ 10\% width	\\
\citet{Manara2015}	&	Line Luminosity	&	Ca~\textsc{ii}	\\
\citet{Manara2016}	&	Continuum Excess	&	3300-7100 \AA	\\
\citet{Manara2017}	&	Continuum Excess	&	3300-7150 \AA	\\
\citet{Manara2020}	&	Continuum Excess	&	3300-7100 \AA	\\
\citet{Manara2021}	&	Line Luminosity	&	H$\alpha$, H$\beta$, He~\textsc{i} $\lambda$\,5876, O~\textsc{i} $\lambda$\,6300	\\
\citet{Mohanty2005}	&	Line Luminosity	&	Ca~\textsc{ii}	\\
\citet{Muzerolle2003}	&	Line Profile, Continuum Excess	&	H$\alpha$, 5500, 6200, 6450, 7100, 8700, 8900 \AA	\\
\citet{Muzerolle2005}	&	Line Profile	&	H$\alpha$	\\
\citet{Natta2004}	&	Line Profile, Line Luminosity	&	H$\alpha$, Pa$\beta$	\\
\citet{Natta2006}	&	Line Luminosity	&	Pa$\beta$, Br$\gamma$	\\
\citet{Nguyen-Thanh2020}	&	Line Profile	&	H$\alpha$ 10\% width	\\
\citet{Petrus2020}	&	Line Luminosity, Line Profile	&	H$\alpha$, H$\alpha$ 10\% width	\\
\citet{Pouilly2020}	&	Line Luminosity	&	H$\alpha$, H$\beta$, Ca~\textsc{ii} $\lambda$\,8542, $\lambda$\,8662	\\
\citet{Rigliaco2011}	&	Continuum Excess, Line Profile	&	U-band photometric excess, H$\alpha$ 10\% width	\\
\citet{Rigliaco2012}	&	Continuum Excess	&	3000--25000 \AA	\\
\citet{Rugel2018}	&	Continuum Excess	&	3000--5500 \AA	\\
\citet{Sallum2015}	&	Line Luminosity	&	H$\alpha$ photometry	\\
\citet{Salyk2013}	&	Line Luminosity	&	Pf$\beta$	\\
\citet{Santamaria-Miranda2018}	&	Line Luminosity, Line Profile	&	H$\alpha$, H 11, Ca~\textsc{ii} line flux, H$\alpha$ 10\% width	\\
\citet{Venuti2019}	&	Continuum Excess,	&	$\approx3400$--3600, $\approx4000$--4750, 3600, 4600, 7100 \AA	\\
	& Line Luminosity		&	H$\alpha$, H$\beta$, H$\gamma$, H$\delta$, H 8, H 9, H 10, H 11,	\\
	&		&	Pa$\beta$, Ca~\textsc{ii} $\lambda$\,3934, He~\textsc{i} $\lambda$\,4026, $\lambda$\,5876	\\
\citet{Wagner2018}	&	Line Luminosity	&	H$\alpha$ photometry	\\
\citet{White2003}	&	Continuum Excess	&	6500 \AA	\\
\citet{Wu2015}	&	Line Luminosity	&	H$\alpha$ photometry	\\
\citet{Wu2017}	&	Line Luminosity	&	H$\alpha$ photometry	\\
\citet{Zhou2014}	&	Continuum Excess	&	3365 \AA\ photometry	\\
\enddata					
\end{deluxetable*}	

\section{Unified Rederivation of Parameters}\label{app:rederivephysical}
\subsection{Physical Parameters}
Below we describe the unified methodology used to determine physical parameters for each object in CASPAR, focusing exclusively on the stars and brown dwarfs.

\textit{Distance}---We assume Gaia EDR3 distances for all objects with measurements in that data release \citep[$\mathrm{N}=599$;][]{Bailer-Jones2021}. If an object was not observed with EDR3, we used its Gaia DR2 distance \citep[$\mathrm{N}=5$;][]{Bailer-Jones2018}.  If the object had no Gaia measurement, we assume the average distance to the region listed in Table~\ref{tab:ages} ($\mathrm{N}=189$ in Appendix~\ref{extra}).  The uncertainty on each Gaia distance is from either \citet{Bailer-Jones2018} or \citet{Bailer-Jones2021}, who estimate the 1$\sigma$ span of the highest density interval on the posterior probability density used to determine the distance.  This 1$\sigma$ span sets the lower and upper bounds of the distance and is not assumed to be symmetric around the median distance.  

\textit{Age}---We assume the age and uncertainty of the star forming region or association as described in Section~\ref{sec:kin_phot_age}.  We refer to the individual papers for full details on age and uncertainty determination.

\textit{Spectral Type}---For objects that have only one measured accretion rate in the literature database ($\mathrm{N}=646$), we assume the literature reference spectral type.  For objects with multiple measured accretion rates ($\mathrm{N}=152$), we assume the most recently measured spectral type.  For objects with no spectral types listed in their reference paper, we searched VizieR \citep{VIZIER2000} and SIMBAD \citep{SIMBAD2000} for a spectral type.  If none was found, but a temperature was given, we either a) calculated the spectral type using the spectral type to temperature conversion if stated in the literature reference, or b) calculated the spectral type using the spectral type to temperature conversions of \citet{Herczeg2014} if there was none stated.
For spectral types without listed uncertainties, we adopt the spectral class uncertainty of \citet{Herczeg2014}: M dwarfs: 0.2 subclass, K8--M0.5: 0.5 subclass, G0--K8: 1 subclass.    

\textit{Effective Temperature}---Temperature is calculated using the spectral type to temperature conversion of \citet{Herczeg2014}. The uncertainty is found by calculating the temperature at the upper and lower limits of the spectral type estimate.  We then take the average value of the difference between the bounds and the given temperature as the uncertainty.

\textit{Mass, Luminosity, Radius, Surface Gravity}---To consistently estimate mass, luminosity and radius, we use the evolutionary models of \citet{Baraffe2015} and the MIST MESA models \citep{Dotter2016, Choi2016, Paxton2011, Paxton2013, Paxton2015}. Using object age and temperature, we interpolate over the isochronal models to determine the mass, luminosity, radius, and surface gravity. We heavily modified the \texttt{isochrone}\footnote{\url{https://github.com/timothydmorton/isochrones}} python package \citep{Morton2015} to work with the \citet{Baraffe2015} models and to interpolate between temperatures (the package currently only interpolates between masses).  After interpolating over age and effective temperature, the best fit mass, luminosity, radius, and surface gravity can be extracted.  To determine the uncertainty on these quantities, we find the lower and upper limits on the age and temperature; this produces four bounds (1-low age/low temperature, 2-low age/high temperature, 3-high age/low temperature, 4-high age/high temperature). We then take the average of the difference between the bound values and the given value as the uncertainty.  
We assume that all reported spectral types and line luminosities have been corrected for extinction.

\subsection{Accretion Parameters}
Below we detail the unified methodology used to derive accretion rates for various accretion diagnostics.

\textit{Continuum Excess}---Accretion Luminosities are determined from the total excess luminosity (derived from spectral template fitting), which has traditionally been assumed to result primarily from the continuum excess.  
From \citet{Herczeg2008}, $L_\mathrm{acc} \propto d^2$.  Therefore, using the original literature accretion luminosities, we scale them by the Gaia distances, and then derive an accretion rate using the updated accretion luminosity, mass, and radius. 

\textit{Line Luminosities}---For papers that report line fluxes (excluding the \citet{Mohanty2005} Ca~\textsc{ii} $\lambda$\,8662 line, see below), we calculate mass accretion rates from the reported line flux under a single set of scaling relations. For references that only report the accretion luminosity or accretion rate of the line(s), we use the $L_\mathrm{line}-L_\mathrm{acc}$ scaling relations and distance given in the paper to infer the measured line flux.  For studies that only give accretion rates, we first calculate $L_\mathrm{acc}$ using the reference mass and radius and then proceed as above.  Once we have derived a line flux, we calculate the accretion rate as follows:
\begin{enumerate}[nosep]
    \item We update the line luminosity using the Gaia DR2 or EDR3 distance and assuming isotropic emission: $L_\mathrm{line} = 4\pi d^2 F_\mathrm{line}$.
    \item We convert line luminosity to accretion luminosity following $\log(L_\mathrm{acc}/L_\odot) = a\times \log(L_\mathrm{line}/L_\odot)+b)$ using a single set of independently derived scaling relations \citep[those of][]{Alcala2017, Salyk2013}.  
    \item We convert to accretion rate using our rederived mass and radius with Equation~(\ref{eqn1}).
\end{enumerate}
For objects with given line flux uncertainties, we propagate this error forward using the uncertainties on Gaia distance, rederived mass and radius uncertainties, and the uncertainties on the scaling relationship.  

The Ca~\textsc{ii} 8662 \AA\ line fluxes calculated by \citet{Mohanty2005} are a unique case, as the line fluxes are calculated assuming a best fit modeled continuum, causing systematic offsets between their values and ``true" values \citep[for full details see][]{Mohanty2005}.  Therefore, we use their scaling relations (their eqns 1. and 3.) in step 2 above.  

\textit{H$\alpha$ Photometric Luminosities}---Narrowband H$\alpha$ photometry has frequently been used to calculate mass accretion rates for substellar objects \citep[e.g.,][]{Kalari2015, Sallum2015, Wu2015, Wu2017, Close2014, Wagner2018}.  Once the H$\alpha$ luminosity is determined, an $L_\mathrm{H\alpha}-L_\mathrm{acc}$ scaling relation can be applied.  Therefore, in order to recalculate the accretion rate, we follow the same procedure as our rederivation of spectroscopic line fluxes (see above), substituting recalculated physical parameters (distance, mass, radius) and adopting a single scaling relation \citep{Alcala2017}.  

However, for 16 objects in the \citet{Kalari2015} Lagoon Nebula sample ($D\sim1326$ pc), Gaia EDR3 distances put them at $\sim 200-700$pc.  \citet{Henderson2012} proposed a that the cluster might be 15\% closer than previous estimates; however, even with a closer distance, these objects are still well below any assumed distance to the cluster. Additionally, ten of the objects have proper motions with Right Ascension and Declination dispersion greater than best-fit values found for the Lagoon Nebula \citep[$\sigma_{\mu_{\alpha}}\sim 4.06$ km/s, $\sigma_{\mu_{\delta}}\sim 2.8$ km/s,][]{Wright2019} assuming a distance of 1326 pc.  
Therefore, we assume these objects are not true members of the Lagoon Nebula, and their true ages are unknown.  Therefore, their masses cannot be estimated properly and we exclude them from further analysis.

\begin{figure}[htp!]
    \centering
    \includegraphics[width=\linewidth]{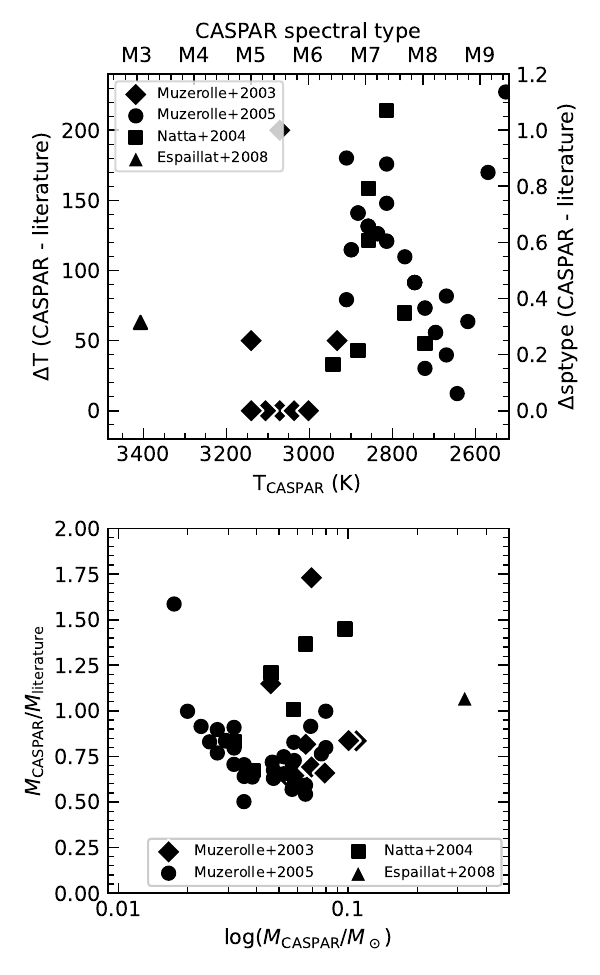}
    \caption{Absolute difference between literature and CASPAR derived temperatures/spectral types as a function of CASPAR temperature/spectral type (\textit{top}) and the ratio between literature and CASPAR derived masses as a function of CASPAR mass (\textit{bottom}) for objects with accretion rates derived from H$\alpha$ emission profile modeling. Estimated accretion rates do not vary with rederived masses and radii if the spectral types are within 1 subclass ($<200$ K difference) and masses are within a factor of two of the original values.  We find all values in CASPAR to be within those parameters.}
    \label{fig4:Haprofiletest}
\end{figure}

\textit{Line Profiles}---Accretion rates found by modeling the H$\alpha$ emission profile with radiative transfer models of the magnetosphere rely on the line-of-sight inclination angle of the disk and velocity field.  Gas velocities are sensitive to mass \citep{Muzerolle2001}. Though the input mass range will vary according to the evolutionary tracks used, the best fit model diverges if the mass is varied by a factor of two or more (due to significant variation in the model gas velocity; $v_\mathrm{gas} \propto \sqrt{M}$).  For object radii, uncertainty results from the spectral type-temperature conversion, with a 200 K error (equivalent to 1 spectral subclass) equal to 30\% error in radius.  Nonetheless, as the modeled gas density is not strongly dependent on radius (is depends on the system geometry), the model is less sensitive to variation.  Therefore, as long as the rederived masses are approximately within factor of two, temperatures within 200 K (1 spectral subclass), and/or radii within 30\% of the literature values, no rederivation is needed. From Fig.~\ref{fig4:Haprofiletest}, we find that the average difference in temperature is 94 K (corresponding to half a spectral class) and the masses are within a factor of 0.83 and therefore the accretion rates do not have to be recalculated. 

Accretion rates have been found to scale directly with H$\alpha$ 10\% width \citep{Natta2004}. Therefore, accretion rates originally found using the \citet{Natta2004} scaling relation do not have to be recalculated; this encompasses all objects with H$\alpha$ 10\% width within CASPAR.

\section{Other linear fitting techniques}\label{app:fittingtechniques}
It is well known that different linear regression procedures should be used based on the data being considered.  Significant work \citep{Isobe1990, Feigelson1992} has explored how these different algorithms affect astronomical data and results.  As we explore the best linear fits for a variety of quantities in CASPAR, reliable fitting is necessary.  \texttt{linmix}, a Bayesian linear regression routine \citep{Kelly2007}, takes into account both $x$ and $y$ measurement errors as well as upper limits. In order to determine the extent to which these quantities affect the linear fits, we compare this fit to other fitting techniques that take into account a variety of these parameters. See Table~\ref{tab:fittingalorithms}. We use \texttt{python} to derive the best fits for each technique:
\begin{itemize}
\item orthogonal distance regression (ODR): \texttt{scipy.ODR}
\item ordinary least squares bisector (BCES): \texttt{bces}
\item weighted least squares (WLS): \texttt{statsmodels.WLS}
\item ordinary least squares (OLS): \texttt{curve\_fit}.
\end{itemize}

\begin{figure}[htp!]
    \centering
    \includegraphics[width=\linewidth]{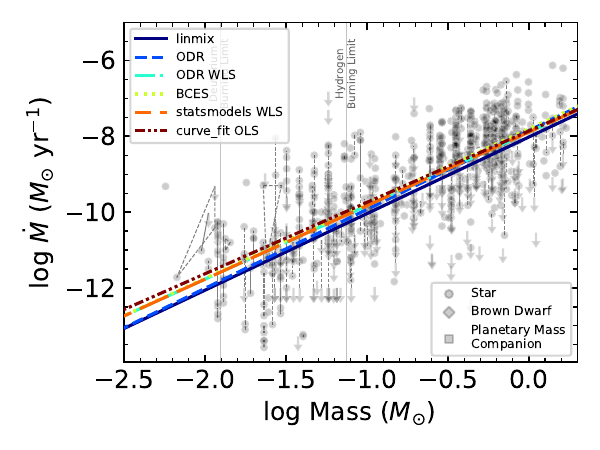}
    \caption{CASPAR Accretion rate vs mass with linear best fits from different linear regression algorithms. \texttt{linmix} takes into account all parameters, while the others take into account some parameters as listed in Table~\ref{tab:fittingalorithms}.}
    \label{fig:appB}
\end{figure}

Fig.~\ref{fig:appB} and Table~\ref{tab:fittingalorithmsresults} show the results. Algorithms which include $x$ errors but do not properly take into account upper limits (ODR, BCES) still closely match \texttt{linmix}, while those that do not take into account errorbars do not reproduce \texttt{linmix}.  This is most prominent in the substellar regime, where the fits diverge significantly. Those without $x$ errors consistently overestimate the intercept of the fit compared to those with take both $x$ and $y$ into account.       

\begin{deluxetable}{lccc}
\tablecaption{Linear regression algorithms\label{tab:fittingalorithms}}
\tablewidth{0pt}
\tablehead{\colhead{fitting algorithm} & \colhead{$y$ err} & \colhead{$x$ err} & \colhead{upper limits}}
\startdata
\texttt{linmix} & x & x & x \\
ODR & x & x & -- \\
ODR WLS & x &-- &--  \\
BCES & x & x & -- \\
WLS & x & -- & -- \\
OLS &-- &-- &-- \\
\enddata
\end{deluxetable}

\begin{deluxetable}{lccc}
\tablecaption{linear regression results \label{tab:fittingalorithmsresults}}
\tablewidth{0pt}
\tablehead{\colhead{fitting algorithm} & \colhead{$a\,(\pm$err)} & \colhead{$b\,(\pm$err)}}
\startdata
\texttt{linmix} & 2.02\,(0.06) & $-8.02\,(0.05)$\\
\texttt{scipy.ODR} & 2.08\,(0.05) & $-7.85\,(0.04)$ \\
\texttt{scipy.ODR} WLS & 1.94\,(0.06) & $-7.89\,(0.05)$ \\
\texttt{bces.BCES} & 1.97\,(0.06) & $-7.80\,(0.05)$ \\
\texttt{statsmodels.WLS} & 1.94\,(0.02) & $-7.89\,(0.02)$ \\ 
\texttt{curve\_fit} OLS & 1.88\,(0.05) & $-7.86\,(0.05)$ \\ 
\enddata
\end{deluxetable}

\begin{deluxetable}{lcccc}
\tablecaption{Population statistics\label{tab:fitcompare}}
\tablewidth{0pt}
\tablehead{\colhead{} & \colhead{mean residual} & \colhead{AIC} & \colhead{$w$} & \colhead{R$^2$} \\
\colhead{} & \colhead{(dex)} & \colhead{} & \colhead{} & \colhead{}}
\startdata
Star fit & 0.12 $\pm$ 0.82 & 1892.66 & 0.99 & 0.35 \\
Best fit & 0.17 $\pm$ 0.82 & 1905.76 & 0.001 & 0.34 \\
\hline 
BD fit   & 0.25 $\pm$ 1.09 & 726.57 & 0.98 & 0.14 \\
Best fit & 0.23 $\pm$ 1.10 & 735.46 & 0.02 & 0.11 \\
\hline 
Planet fit & 0.05 $\pm$ 0.98 & 110.73 & 0.99   & 0.002 \\
Best fit   & 0.82 $\pm$ 0.99 & 131.25 & $4\times10^{-5}$ &  $-0.71$ \\
\enddata
\tablecomments{AIC: Akaike Information Criterion, $w$: Akaike weights, R$^2$: coefficient of determination }
\end{deluxetable}

\begin{figure*}[htp!]
    \centering
    \includegraphics[width=\linewidth]{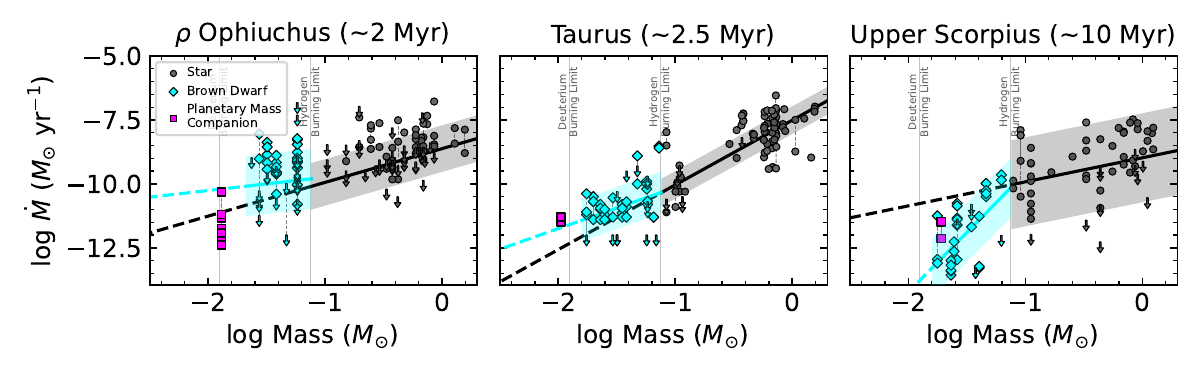}
    \caption{Accretion rate vs mass for $\rho$ Ophiuchus (left), Taurus (center), and Upper Scorpius (right).  Colors and markers are as in Fig.~\ref{fig:MMdot_massregions}.  The best linear fits (solid line) and $1\sigma$ scatter (band) are shown for the star (black) and brown dwarf (cyan) populations.  The dashed lines show the extrapolation into the planetary regime.}
    \label{fig:SFRfits}
\end{figure*}

\section{Separate Mass Population Fit statistics} \label{app:popstats}
In Table~\ref{tab:fitcompare}, we compare the mean residual from each fit, the AIC, and Akaike weights ($w$) for each fit.  As our models have the same number of parameters, the AIC will inform the goodness of fit between the two models, with the minimum AIC corresponding to the preferred model.  The Akaike weights are the relative likelihoods of the models; we assume that $w>0.95$ indicates the statistically favored model.  

\textit{Stars:} We find similar AIC statistics for the best total fit to CASPAR and star-only fit, with neither model preferentially preferred. This indicates that either fit can be used to model the data. 

\textit{Brown Dwarfs:} In the substellar regime, we find a small, though  not statistically significant, difference in the residuals between the CASPAR total fit and the brown dwarf-only fit, with the mean decreasing from 0.25 to 0.23.  However, from the AIC statistics and weights shown in Table~\ref{tab:fitcompare}, we can show that the BD-only fit is more significantly preferred than the CASPAR total best fit. 

\textit{PMCs:} The best fit clearly showed non-normal residuals for the PMCs.  The PMC-only best fit is significantly preferred with the mean of the residuals decreasing from 0.82 to 0.05.

\section{Separate Mass Population fits for Individual Star Forming regions}\label{app:SFR}
We examine substellar and stellar population best fits for individual star forming regions in CASPAR, removing the need to fit for age.  We select the $\rho$ Ophiuchus ($\sim2$~Myr), Taurus ($\sim2.5$~Myr), and Upper Scorpius ($\sim10$~Myr) regions as they span a wide range of ages.  As shown in Fig.~\ref{fig:SFRfits}, from just these three regions, we see a steepening slope ($\alpha=0.52, 1.61, 4.66$) with age in the substellar regime, while the stellar regime appears to slightly flatten ($\alpha=1.32, 2.53, 0.94$) for $\rho$ Ophiuchus, Taurus, and Upper Scorpius, respectively, similar to the trends seen by \citet{Alcala2017} and \citet{Manara2017b}.

\bibliography{ref}
\bibliographystyle{yahapj.bst}

\end{document}